\documentclass[journal,10pt]{IEEEtran}
\ifCLASSINFOpdf
\else
\fi
\usepackage{enumitem}
\usepackage{caption}
\usepackage{algpseudocode}
\usepackage{mathtools}
\usepackage{algorithm}
\usepackage{setspace}
\usepackage{graphicx}
\usepackage{notoccite}
\usepackage{makecell}
\usepackage{amssymb}
\usepackage{url}
\usepackage[table]{xcolor}
\usepackage{placeins}
\usepackage{amsfonts}
\usepackage{amsthm}
\usepackage{amsmath}
\usepackage{bbm,balance}
\usepackage{subfigure}
\usepackage{multirow}
\usepackage{pdfrender}
\usepackage{color}
\usepackage{calc}
\usepackage{booktabs}
\usepackage{multirow}
\usepackage{siunitx}
\usepackage{fancybox}
\usepackage{array}

\usepackage{mathtools}

\begin{document}

\title{Dynamics of Laser-Charged UAVs: A Battery Perspective}

\author{Wael Jaafar,~\IEEEmembership{Senior Member,~IEEE,}
        and Halim Yanikomeroglu,~\IEEEmembership{Fellow,~IEEE}
 
\thanks{W. Jaafar and H. Yanikomeroglu are with the Department of Systems and Computer Engineering, Carleton University, Ottawa,
ON, Canada (e-mails: \{waeljaafar, halim\}@sce.carleton.ca).}
\thanks{This  work  is  supported  in  part  by  the  Natural  Sciences  and Engineering Research Council Canada (NSERC) and in part by Huawei Canada.}
\thanks{Copyright (c) 2020 IEEE. Personal use of this material is permitted. However, permission to use this material for any other purposes must be obtained from the IEEE by sending a request to pubs-permissions@ieee.org.}
}

%\author{Author 1,
%        Author 2
 
%\thanks{Author 1 and Author 2 are with Affiliation.}

\markboth{IEEE Internet of Things Journal,~Vol.~XX, No.~XX, Month~Year}%
{Shell \MakeLowercase{\textit{et al.}}: Bare Demo of IEEEtran.cls for IEEE Journals}

\maketitle

\begin{abstract}
In this paper, we
{aim to sustain unmanned aerial vehicle (UAV) based missions for longer periods of times through different techniques. First, we consider on-the-mission UAV recharging by a low-power laser source (below 1 kilowatt). In order to achieve the maximal energy gain from the low-power laser source, we propose an operational compromise, which consists of the UAV resting over buildings with cleared line-of-sight to the laser source. Second, to provide a precise energy consumption/harvesting estimation at the UAV, we} investigate the {latter's dynamics in a mission environment. Indeed, we}
%at achieving a mission while conserving as much energy as possible. 
study the UAV's battery dynamics by leveraging the electrical models for motors and battery. Subsequently, using these models, the path planning problem in a particular Internet-of-Things based use-case is revisited from the battery perspective. {The objective is to extend the UAV's operation time using both laser-charging and accurate battery level estimation.} {Through} a graph theory approach, the problem is solved optimally, and compared to benchmark trajectory approaches. {Numerical results demonstrate} the efficiency of this novel {battery}
perspective for all path planning approaches. In contrast, we found that the energy perspective is very conservative and does not exploit optimally the available energy resources. {Nevertheless, we propose a simple adjustment method to correct the energy perspective, by} carefully evaluating the energy as a function of the UAV motion {regimes}.
%A heuristic solution is proposed, then implemented in order to prove its efficiency. 
Finally, the impact of {several} parameters, such as turbulence and distance to charging source, is studied.
%The objective is to provide a simple yet practical model of quadrotor UAV consumed/harvested energy and battery dynamics for researchers conducting work on energy-efficient aerial networks.  
\end{abstract}



\IEEEpeerreviewmaketitle

\vspace{-10pt}
\section{Introduction}
Unmanned aerial vehicles (UAVs) have been experiencing a boom in interest lately from industry and research. Indeed, several new applications that rely on UAVs have emerged in recent years in connection with the evolution of wireless networks into 5G and beyond \textcolor{black}{\cite{Bor2016}}. UAVs have been deployed for aerial based applications such as security inspection \cite{Liu2019}, precision agriculture \cite{Tokekar2016}, traffic control \cite{Zhu2018_2}, and package delivery \cite{Sawad2019}. They can also act as cellular base-stations to provide connectivity to rural and disaster-hit areas. \textcolor{black}{For instance, authors in \cite{Kaleem2019} proposed a three-layer architecture, in which an edge-cloud layer is formed temporarily by UAVs to provide edge computing and connectivity during disaster instances, while in \cite{Yan2019}, the authors investigated the joint UAV access selection and base station (BS) bandwidth allocation problem of a UAV-assisted Internet-of-Things (IoT) communication network.} Hence, \textcolor{black}{UAVs} are seen as a promising technology to profit businesses and help society.

\subsection{{Related Work}}
Despite all their promise, most battery-powered UAVs have a major drawback: their flight duration is significantly limited and thus are unable to satisfy the requirements of all these emerging applications. This limitation is mainly due to the limited on-board lithium-ion polymer (LiPo) battery capacity. 
To overcome this limitation, several works focused on reducing on-board energy consumption through UAV or battery hot-swapping \cite{Galkin2018UAVsAM}, battery capacity increase \cite{Saha2011}, UAV placement or path optimization \cite{Alzenad2018,Zeng2017}. 
Unfortunately, these approaches are unable to deliver a significant flight time increase and satisfy the requirements of UAV-based applications. \textcolor{black}{Recently, powering through laser beaming has been proposed to enable longer UAV flight times \cite{Diamant2018}\nocite{Ouyang2018,Zhang2018,Pan2019}--\cite{Lahmeri2019}. Its potential has encouraged several companies to develop this technology \cite{Nugent2010,powerlight}. Laser beaming can be implemented using a laser array oriented through an optical system (e.g., set of mirrors or diamonds) towards the collecting lens of a targeted UAV. Alternatively, distributed laser charging (DLC) advocates the use of photo-voltaic cells instead of the collecting lens at the receiver, which is cost-effective and practical in small UAVs \cite{Zhang2018}.   
For efficient charging, a line-of-sight (LoS) link between the charging source and UAV is required. It is to be noted that due to current technology limitations, UAVs cannot rely solely on laser charging; besides, LoS (which is necessary for laser charging) cannot be guaranteed all the time. Also, equipping the UAV with more battery cells would make the UAV heavier, which is counterproductive as the target is to extend the UAV’s operating time. Complementing a lighter UAV (having a small battery) with laser charging can be seen as a hybrid approach, as it would allow operating the UAVs for extended times.}

%Although interesting preliminary results were obtained, laser beaming is still impractical due to its high cost and its limitation to beaming towards a single UAV at a time.   
%In this context, non-electromagnetic field charging has been emerged as a promising wireless power transfer (WPT) technology to charge UAVs \cite{Zhang2018}. Indeed, using photo-voltaic cells, distributed laser charging (DLC) can be adopted, where energy is captured by any UAV via a line-of-sight (LoS) link.

On the other hand, the importance of energy consumption and harvesting have received a limited consideration from the battery perspective. Indeed, almost all works investigating UAV-based issues, such as 3D trajectory optimization and resource allocation, focused on the energy perspective of consumed/harvested average or instantaneous power when flying, hovering, data processing, or communicating, without taking into account the relation to the initial available energy, and impact of turbulence forces.  
In \cite{Zeng2017}, the authors derived a theoretical model for propulsion energy consumption of fixed-wing UAVs, expressed as a function of the UAV’s flying speed and acceleration. Then, they studied the energy-efficiency (EE) maximization problem subject to the UAV’s trajectory constraints, including departure and arrival locations, minimum and maximum speeds, and maximum acceleration. The EE was defined as the ratio of the total amount of information that can be transmitted from the UAV to the consumed propulsion energy, during an observation time period. The obtained results show a significant EE gain over baseline approaches. A similar problem is investigated by Zeng \textit{et al.} in \cite{zeng2018energy} for a rotary-wing UAV, where the latter was deployed to serve IoT ground nodes. The objective was to minimize the total UAV consumed propulsion and communication energy while satisfying a predefined data rate requirement. The energy minimization problem is formulated, where the UAV trajectory and communication time allocation among IoT nodes are jointly optimized, as well as the mission completion time. First, by following the fly-hover-communicate design, the problem is solved by leveraging the traveling salesman problem with neighborhood and convex optimization approaches. Then, the general case, where the UAV communicates while flying, is solved sub-optimally through successive convex optimization. Numerical results illustrate the superiority of the proposed solutions compared to benchmark schemes.  
In contrast, authors of \cite{Chakareski2019} studied the downlink transmission of a multi-band heterogeneous network, where UAVs can be deployed as small BSs. 
They formulated a two-layer optimization problem to maximize the EE of the system, where  
both the coverage radius of UAVs and radio resource allocation are optimized subject to minimum quality-of-service (QoS) and maximum transmit power constraints. EE was defined as the ratio of the aggregate user data rate delivered by the system to its aggregate energy consumption, which is limited to downlink transmission power and circuitry power. Presented results demonstrate the potential of a UAV tier in the heterogeneous network, which can nearly double the system's EE for particular QoS requirements.
Also, the authors of \cite{Khami2019} optimized jointly the trajectory, speed, and acceleration of a UAV user, aiming to minimize its propulsion power while travelling between two locations. Although non-convex, the authors reformulated this problem to a more tractable form, and solved it using an iterative successive convex approximation technique.  
In a green energy context, Sun \textit{et al.} studied in \cite{Sun2019} joint trajectory and wireless resource allocation for solar-powered UAVs, aiming to maximize ground users' sum throughput over a given time period. 
The aerodynamic power consumption, solar energy harvesting, a finite energy storage capacity, and the QoS requirements of ground users were taken into account in the problem formulation. Optimal UAV trajectory, power, and subcarrier allocation are obtained through monotonic optimization in the offline case. Then, the online problem, where only real-time and statistical knowledge of the channel gains are available, is solved using a low-complexity iterative sub-optimal scheme, which is based on successive convex approximation. The results revealed the near-optimality of the proposed online schemes. Also, a tradeoff between solar energy harvesting and power-efficient communication is identified, where solar energy harvesting is preferred at high altitudes, before moving to lower altitudes to reliably serve ground users. 
Finally, Sekander \textit{et al.} presented in \cite{sek2019performance} novel models of energy harvesting from renewable energy sources, namely solar, wind, and hybrid solar and wind. They derived closed-form expressions of energy-outage probability at harvesting UAVs and SNR outage at ground users for both solar and wind harvesting scenarios. Analytical results were validated through simulations, exhibiting insights on the optimal UAV flight time and transmit power as functions of the harvested energy.

\subsection{{Contributions}}

Although interesting, the aforementioned works did not establish a clear link between the UAV motion model, its energy consumption/harvesting, and battery capacity. For instance, \cite{Zeng2017,zeng2018energy}--\cite{sek2019performance} considered a motion model that ignored the turbulence's effect, while \cite{Chakareski2019} simply neglected the propulsion power. Moreover, \cite{Sun2019}--\cite{sek2019performance} assumed quasi-static flight equilibrium conditions, hence nulling the UAV speed and acceleration effects. Finally, the few works that have discussed the battery capacity \cite{zeng2018energy,Khami2019}--\cite{Sun2019} were limited to a conventional energy perspective, where the effect of the UAV rotor's operating voltage on the real amount of consumed energy is ignored. 
Indeed, a battery perspective is of capital importance since battery-powered UAVs suffer from uncertain estimation of charge level, and hence most mission plans are highly-conservative. In fact, UAV batteries may be affected by the storage state-of-charge (SOC), imposed discharge or load profile, and the variable requirements of flight regimes (takeoff/landing/travelling/hovering). These factors might degrade the terminal voltage that defines the battery shut-off criteria, as the battery SOC nears empty \cite{Saha2011}. 
Hence, we establish here the energy-battery relationship, and show its relevance when investigating a UAV-based challenge such as path planning. {Moreover, efficient on-the-fly laser recharging requires a kilowatts (kW) transmitter, as well as optimized positioning and/or tracking, to gain an acceptable amount of energy at the UAV. However, such a transmitter can be harmful to the surrounding environment \cite{Hassan2020}. Hence, we envision here the use of a low-power laser source (below 1 kW) that does not require specific positioning or tracking. In this work, to efficiently use low-power laser recharging, we propose an operational compromise consisting of the UAV resting over buildings (when available) with cleared LoS to the laser source.} To the best of our knowledge, this is the first work that revisits a UAV-based issue from the battery perspective, while taking into account on-the-mission {low-power} laser-charging.  
The main contributions of this paper may be summarized as follows: 
\begin{enumerate}
    \item {We establish for the first time the relationship between the UAV motion regimes, energy consumption/harvesting, and battery dynamics.}
    
    %First, we expose the {motion and} energy {models} of an on-the-mission laser-charged UAV, and establish {their} relation to the battery dynamics. 
    \item To prove the relevance of the battery perspective in mission design, a UAV path planning problem is revisited, {where the objective is to sustain the mission for the longest period of time with the help of laser charging and accurate battery level estimation. The formulated problem is} solved optimally, {then} compared to different benchmark trajectory approaches. It is found that the battery perspective outperforms the energy perspective, and the latter is very conservative for mission design. 
    
    \item {We propose a practical UAV operational modification to achieve high gains from a low-power (below 1 kW) laser recharging source.}
    
    \item {We achieve a simple, yet precise, adjustment approach to the} energy perspective, {through} carefully evaluating the consumed{/harvested} energy for different {UAV} motion regimes, {e.g.}, hovering, flying, etc. 
    \item Finally, the impact of parameters, e.g., {desired SOC level,} turbulence, and distance to charging source, is investigated, {from the battery perspective.}
    %showing the different influence of lateral and vertical turbulence, as well as the importance of the distance between the UAV and laser source for efficient recharging.  
\end{enumerate}
%To prove the relevance of the battery \textcolor{black}{perspective in mission planning}, a UAV path planning problem is revisited %from the energy/battery perspective 
%and solved optimally \textcolor{black}{and compared to different benchmark trajectory approaches}. 

The remaining of the paper is organized as follows. In Section II, the UAV energy model is presented. Section III details the associated battery dynamics, while section IV formulates the revisited path planning problem and exposes the \textcolor{black}{solution approach}. Section V presents the numerical results. Finally, Section VI concludes the paper.   

%\begin{figure}[t]
%	% Requires \usepackage{graphicx}
%	\centering
%	\includegraphics[width=125pt]{Quadrotor_UAV.pdf}
%	\caption{Quadrotor UAV.}
%	\label{Fig:UAV}
%\end{figure}

%\vspace{-15pt}
\section{UAV Energy Model}
In this section, we provide the expressions of consumed/harvested energy by a quadrotor UAV. Consumed energy is defined by
$E_{c}=E_{\rm{trav}}+E_{\rm{hov}}+E_{\rm{comm}}$, where $E_{\rm{trav}}$ is the energy to travel between locations, $E_{\rm{hov}}$ is the hovering energy, and $E_{\rm{comm}}$ is the communication energy.

For motion control, we consider the model of \cite{Mozaffari2019}, where by adequately adjusting the rotors velocities $v_r$ ($r=1,\ldots,4$), the UAV can hover or travel vertically or horizontally. This model assumes that a path from location $[0,0,0]$ to 3D-destination $\mathbf{w}_D$ (assuming no obstacles along the path) can be broken into six stages, \textcolor{black}{as detailed in (eqs.(44)--(49), \cite{Mozaffari2019})}. In contrast to the simplified motion model presented in \textcolor{black}{\cite{Zeng2017} and \cite{zeng2018energy}}, the model of \cite{Mozaffari2019} captures both the instantaneous \textcolor{black}{3D} UAV movement flexibility and impact of turbulence. Consequently, the motion control energy consumed by the UAV between times $t_0$ and $t_f$ can be given by \cite{Cord1977}
\begin{equation}
%\small
\label{eq:energy}
E= \int_{t_0}^{t_f} \sum_{r=1}^4 e_r(t)i_r(t) \; dt,
\end{equation}
where $e_r(t)$ and $i_r(t)$ are the voltage and current across motor $r$ respectively. 
In steady-state conditions, they are written \cite{Morbidi2016}
\begin{equation}
%\small
    \label{eq:DCmotor}
    e_r(t)= R i_r(t) + \kappa_E v_r(t),
\end{equation}
and
\begin{equation}
%\small
    \label{eq:current}
    i_r(t)=\frac{1}{\kappa_T}\left[ T_f + \kappa_0 v_r^2(t)+D_f v_r(t)+ J \frac{\partial v_r(t)}{\partial t} \right],
\end{equation}
where $R$ is the resistance, $\kappa_E$ is the motor's voltage constant, $\kappa_T$ is the torque constant, $T_f$ is the motor friction torque, $\kappa_0$ is the drag coefficient, $D_f$ is the motor viscous damping coefficient, and $J$ is the rotor inertia.
By combining (\ref{eq:DCmotor})-(\ref{eq:current}) into (\ref{eq:energy}), the latter can be written
\begin{eqnarray}
%\small
\label{eq:energy2}
E &=& \int_{t_0}^{t_f} \sum_{r=1}^4 \Big( \sum_{i=0}^4  c_{i+1} v_r(t)^i  +  \frac{\partial v_r(t)}{\partial t} \big[ c_6 + c_7 \frac{\partial v_r(t)}{\partial t}\nonumber \\
&+&c_8 v_r(t) + c_9 v_r(t)^2 \big] \Big) \; dt,
\end{eqnarray}
where $c_1, \ldots, c_9$ are expressed as
%\small
\begin{eqnarray}
\label{constants}
&&c_1=\frac{R T_f^2}{\kappa_T^2}, \; c_2= \frac{T_f}{\kappa_T}\left( \kappa_E + \frac{2 R D_f}{\kappa_T} \right), \; c_4= \frac{\kappa_0}{T_f} c_2,\nonumber 
%&&c_3=\frac{D_f}{\kappa_T}\left( \frac{R D_f}{\kappa_T}+ \kappa_E \right) + \frac{2 R T_f \kappa_0}{\kappa_T^2}, \; c_5=\frac{\kappa_0^2}{T_f^2} c_1,\nonumber 
%&&c_6 = \frac{2 J}{T_f}c_1, \; c_7=\frac{J^2}{T_f^2} c_1, \; c_8=\frac{J}{T_f}c_2, \; c_9=\frac{\kappa_T}{T_f}c_6. \nonumber
\end{eqnarray}
\begin{eqnarray}
%\label{constants}
%&&c_1=\frac{R T_f^2}{\kappa_T^2}, \; c_2= \frac{T_f}{\kappa_T}\left( \kappa_E + \frac{2 R D_f}{\kappa_T} \right), \; c_4= \frac{\kappa_0}{T_f} c_2,\nonumber \\
&&c_3=\frac{D_f}{\kappa_T}\left( \frac{R D_f}{\kappa_T}+ \kappa_E \right) + \frac{2 R T_f \kappa_0}{\kappa_T^2}, \; c_5=\frac{\kappa_0^2}{T_f^2} c_1,\nonumber \\
&&c_6 = \frac{2 J}{T_f}c_1, \; c_7=\frac{J^2}{T_f^2} c_1, \; c_8=\frac{J}{T_f}c_2, \; c_9=\frac{\kappa_T}{T_f}c_6. \nonumber
\end{eqnarray}
\noindent
\normalsize
Consequently, travelling energy $E_{\rm{trav}}$ can be written as
\begin{equation}
    %\small
    \label{eq:Etrav}
    E_{\rm{trav}}=\sum_{s=1}^5 E_{s},
\end{equation}
where $E_{s}$ is the consumed energy in stage $s$ ($s = 1, \ldots, 5$), \textcolor{black}{where $s=1,3,5$ correspond to orientation change, and $s=2,4$ to displacement stages.} 
Since $v_r$ are constant (eqs. (44)--(48), \cite{Mozaffari2019}), the consumed energy can be given by
\begin{eqnarray}
%\small
\label{eq:Es1}
&&E_{s}=\left(\tau_{(2.5s+1.5)} - \tau_{(2.5s-2.5)}\right) \Big( 3 c_1 + (1+\sqrt{2})c_2 v_{\rm{max}}\nonumber \\
&&+ 2 c_3 v_{\rm{max}}^2 + (1+\frac{1}{\sqrt{2}})c_4 v_{\rm{max}}^3 + \frac{3}{2} c_5 v_{\rm{max}}^4 \Big),\; s=1,3,5, \nonumber
%&&E_{s}= \left(\tau_{2.5s}-\tau_{(2.5s-1)}\right) \cdot 4 \sum_{i=1}^5 c_i v_{\rm{max}}^{i-1},\; s=2,4,
\end{eqnarray}
\begin{eqnarray}
%\small
%\label{eq:Es1}
%&&E_{s}=\left(\tau_{(2.5s+1.5)} - \tau_{(2.5s-2.5)}\right) \Big( 3 c_1 + (1+\sqrt{2})c_2 v_{\rm{max}}\nonumber \\
%&&+ 2 c_3 v_{\rm{max}}^2 + (1+\frac{1}{\sqrt{2}})c_4 v_{\rm{max}}^3 + \frac{3}{2} c_5 v_{\rm{max}}^4 \Big),\; s=1,3,5, \nonumber \\
&&E_{s}= \left(\tau_{2.5s}-\tau_{(2.5s-1)}\right) \cdot 4 \sum_{i=1}^5 c_i v_{\rm{max}}^{i-1},\; s=2,4,
\end{eqnarray}
\noindent
\normalsize
where $v_{\rm{max}}$ is the maximal velocity and $\tau_j$ ($\tau_1<\ldots<\tau_{14}$) are the switching times at which UAV control inputs change, given in (Appendix D, \cite{Mozaffari2019}).

In the presence of an external force, e.g., gravity and turbulence, the UAV's hovering energy can be written as
\begin{equation}
E_{\rm{hov}}= \Delta \cdot 4 \sum_{i=1}^5 c_i \left({\frac{|\textbf{F}_e|}{4 \varrho}}\right)^{\frac{i-1}{2}},
\end{equation}
where $\Delta$ is the hovering duration, $|\mathbf{F}_e|$ is the external force amplitude, and $\varrho$ is the lift coefficient, whereas, the communication energy of the UAV is expressed by 
\begin{equation}
    %\small
    \label{eq:Ecomm}
    E_{\rm{comm}}= \int_{t_0}^{t_f} \sum_{u=1}^U P_u(t) dt,
\end{equation}
where $P_u(t)$ is the communication power to node $u$ in time $t$, and $U$ is the total number of nodes.
Finally, \textcolor{black}{the received charging energy from a DLC source is given by \cite{Zhang2018}}
\begin{eqnarray}
\label{eq:EEH}
%\small
E_{\rm{harv}}&=& \int_{t_0}^{t_f} P_0(t) dt=a_1 a_2 \int_{t_0}^{t_f} \nu(t)P_s(t) dt\nonumber \\  
%E_{\rm{harv}}=\int_{t_0}^{t_f} P_0(t) dt= (t_f-t_0) \frac{a P_s e^{-\alpha d}}{D+ d \Delta \theta},
&+& a_2 b_1 \int_{t_0}^{t_f} \nu(t) dt+ b_2 \left(t_f - t_0\right),
\end{eqnarray}
%\begin{eqnarray}
%\small
%%E_{\rm{harv}}&=& a_1 a_2 \int_{t_0}^{t_f} \nu(t)P_s(t) dt\nonumber \\  
%&+& a_2 b_1 \int_{t_0}^{t_f} \nu(t) dt+ b_2 \left(t_f - t_0\right),
%\end{eqnarray}
%where $P_0(t)$ is the received power, $P_s$ is the laser source transmit power (assumed constant), $\Delta \theta$ is the angular spread, $\alpha$ is the attenuation coefficient, $D$ is the size of the initial beam, and $a=a_1 a_2 a_3$, with $a_1$ is the laser energy harvesting efficiency, $a_2$ is the receiver's collecting lens area, and $a_3$ is the transmit receiver optical efficiency. Expression (\ref{eq:EEH}) is valid when charging occurs for a resting UAV (i.e., at a fixed location during charging). It is to be noted that laser charging and communication operations occur on different frequency bands (e.g., above 1 THz for laser charging \cite{Lahmeri2019} and below 6 GHz for communication \cite{Alzenad2018}), hence the UAV-user wireless link does not experience interference from laser charging.} 
where $P_0(t)$ (resp. $P_s$) is the received (emitted) power, $a_i$ and $b_i$ are curve fitting parameters, $\nu(t)=e^{{-\alpha d}}$ is the average transmission efficiency, $d$ is the distance between the UAV and DLC source \textcolor{black}{at time $t$}, and $\alpha$ is the laser attenuation coefficient \cite{Zhang2018}. \textcolor{black}{It is to be noted that laser charging and communication operations occur on different frequency bands (e.g., above 1 THz for laser charging \cite{Lahmeri2019} and below 6 GHz for communication \cite{Alzenad2018}), hence the UAV-user wireless link does not experience interference from laser charging.}

%\vspace{-5pt}
\section{Kinetic Battery Model and Dynamics}
\subsection{Background}
Due to the different flight regimes of UAVs, and variable load and discharge profiles, it is important to present an accurate relationship between energy and battery dynamics. In the literature, besides the theoretical Peukert's law that describes the discharge profile of a battery \cite{Linden1995}, several battery models have been proposed for different applications. For instance, authors of \cite{Doyle1993} characterized the battery electro-chemically using six differential equations, while the authors in \cite{Hageman1993} proposed an electrical circuit-based model. Despite their accurate battery characterization, these models are complex to manage in a performance-oriented setup. Later, preference has been shifted towards the Kinetic Battery Model (KiBaM) \cite{MANWELL1993} and the diffusion-based model \cite{Rakhmatov2001AnAH}. With only two differential equations to fully describe the battery behavior, they are seen as low-complex and practical models \cite{Jongerden2009}.

\subsection{UAV Battery Model and Dynamics}
Since most UAVs use LiPo batteries and this type of batteries can be well modeled using KiBaM, we opt here for this model \cite{MANWELL1993,Jongerden2009} and \cite{Jongerden2017}. 
\begin{figure}[t]
\includegraphics[width=0.99\linewidth]{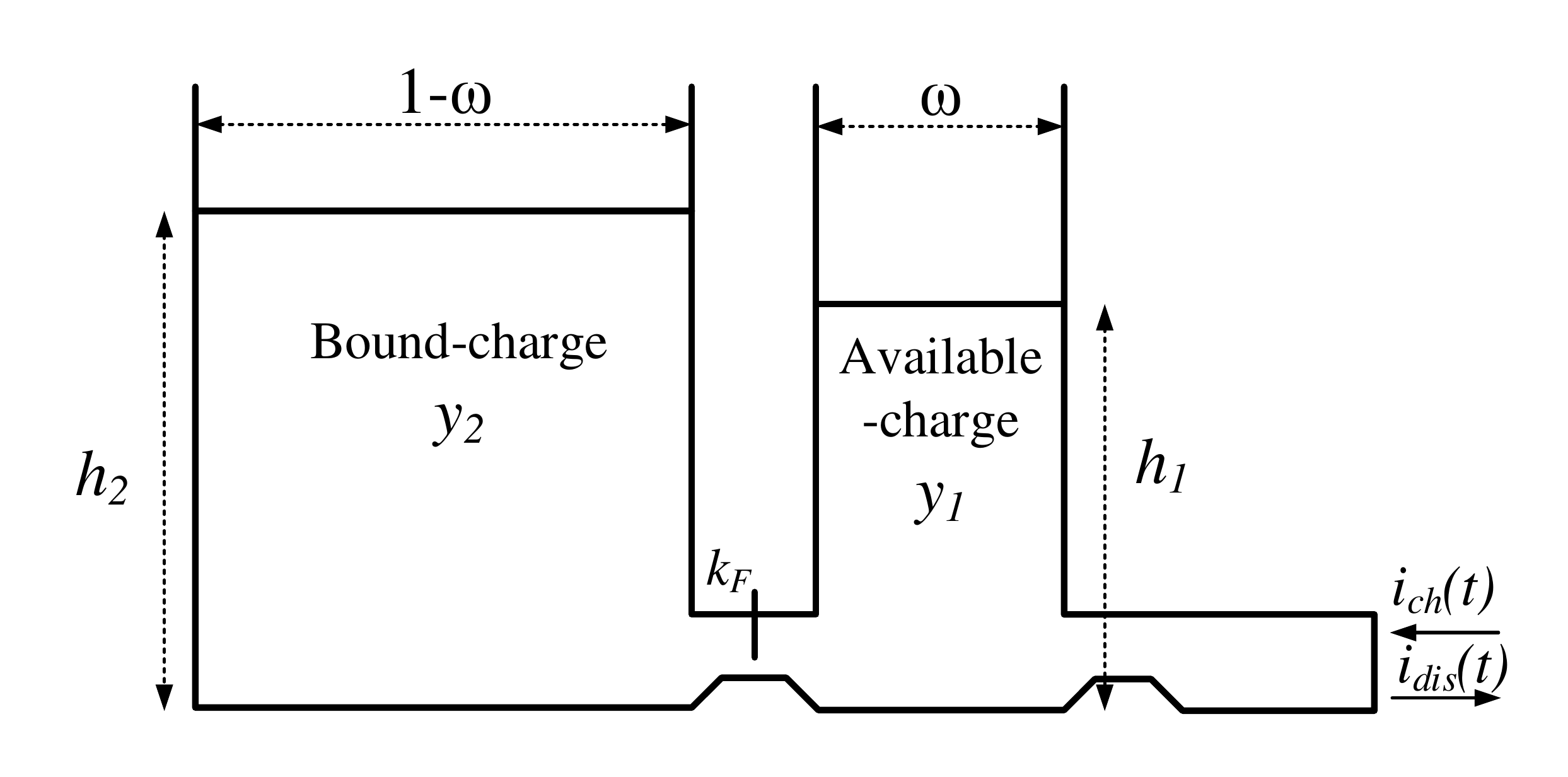}
   \caption{{The KiBaM battery model.}}
    \label{fig:kibam}
\end{figure}

{As illustrated in Fig. \ref{fig:kibam}, in} KiBaM, the battery charge is divided into two wells: an available-charge well ($y_1$) and a bound-charge well ($y_2$). Given $t \in [t_0,t_f]$, and the initial battery conditions $y_1(t_0)=\omega B$ and $y_2(t_0)=(1-\omega)B$, where $B$ is the battery capacity and $\omega \in [0,1]$ is the splitting factor of well levels, the change in charge of both wells is described by \cite{Jongerden2017}:
%\begin{subequations}
\begin{eqnarray}
%\small
    \label{eq:well1}
    \frac{\partial y_1 (t)}{\partial t} &=& \bar{i}(t) + k_F \left( h_2(t)-h_1(t) \right)%\\
    %\label{eq:well2}
    %\frac{\partial y_2 (t)}{\partial t}&=& -k_F \left( h_2(t)-h_1(t) \right),
\end{eqnarray}    
%\end{subequations}
\begin{eqnarray}
%\small
    %\label{eq:well1}
    %\frac{\partial y_1 (t)}{\partial t} &=& \bar{i}(t) + k_F \left( h_2(t)-h_1(t) \right)\\
    \label{eq:well2}
    \frac{\partial y_2 (t)}{\partial t}&=& -k_F \left( h_2(t)-h_1(t) \right),
\end{eqnarray}    
where $k_F$ controls the flowing rate between the wells, $h_1(t)=y_1(t)/\omega$ and $h_2(t)=y_2(t)/\left(1-\omega\right)$ are the heights of the wells, and
\begin{equation}
%\small
\bar{i}(t) = \left\{
    \begin{array}{ll}
         i_{\rm{ch}}(t) & \mbox{in the charge state}\\
        -i_{\rm{dis}}(t) & \mbox{in the discharge state},
    \end{array}
\right.
\end{equation}
where $i_{\rm{ch}}(t)$ and $i_{\rm{dis}}(t)$ are the recharge and discharge currents of the UAV's battery respectively. On one hand, we assume KiBaM constant current charging, where $i_{\rm{ch}}(t)=I_{\rm{ch}}$\footnote{Usually, charging has two phases, the first at constant maximum current until maximum voltage is reached, and the second at constant maximum voltage to keep the level of the available charge well at its maximum \cite{Jongerden2017}. Since current \textcolor{black}{wireless recharging} technologies cannot recharge a flying UAV fully, only the first phase can be achieved.}. To extend the battery life, it is recommended that $I_{\rm{ch}}$ should not exceed 1C$\times B$, where 1C is a measure of the charge current, known as C-rating, and $B$ value in Ah. 
Given the nominal voltage of the LiPo battery $e_{\rm{nom}}$, 
%\textcolor{black}{this means $P_0 \leq I_{\rm{ch}}e_{\rm{nom}}$, or equivalently $P_s \leq \frac{I_{\rm{ch}}e_{\rm{nom}}}{C}(D+d \Delta \theta)e^{\alpha d}$.}
we have the constraint $P_0 \leq I_{\rm{ch}}e_{\rm{nom}}$, i.e., the DLC source power should respect 
\begin{equation}
 P_s(t) \leq \frac{I_{\rm{ch}} e_{\rm{nom}} - a_2 b_1 \textcolor{black}{\nu(t)} - b_2}{a_1 a_2 \textcolor{black}{\nu(t)}}.   
\end{equation}
%%harvested power respects $P_0(t) \leq P_{\rm{ch}}=I_{\rm{ch}} \times e_{\rm{nom}}$. Using (\ref{eq:EEH}), we obtain $P_s(t) \leq \frac{P_{\rm{ch}} - a_2 b_1 \nu - b_2}{a_1 a_2 \nu}$. 
On the other hand, $i_{\rm{dis}}(t)=i_{\rm{cont}}(t)+i_{\rm{comm}}(t)$, where $i_{\rm{cont}}(t)=\sum_{r=1}^4 i_r(t)$ is the UAV's control current, obtained using (\ref{eq:current}), and $i_{\rm{comm}}(t)=\frac{P_U}{e_{\rm{tr}}}$ is the communication current, where $e_{\rm{tr}}$ is the UAV transceiver's voltage.
By solving (\ref{eq:well1})--(\ref{eq:well2}) for constant $\bar{i}(t)=\Bar{I}$, we obtain the battery levels at $t_f$ \cite{Jongerden2009}
%\begin{subequations}
\begin{eqnarray}
%\small
\label{eq:y1}
y_1(t_f)&=& y_1(t_0) e^{-k' \delta } + \frac{\left( y(t_0) k' \omega + \bar{I} \right) \left( 1 - e^{-k' \delta } \right) }{k'} \nonumber \\
&+& \frac{\bar{I} \omega \left( k' \delta  -1 + e^{-k' \delta } \right) }{k'},
%\label{eq:y2}
%y_2(t_f)&=& y_2(t_0)e^{-k' \delta } + y(t_0)(1-\omega)\left(1-e^{-k' \delta }\right)\nonumber \\
%&+& \frac{\bar{I} (1-\omega) \left( k' \delta  -1+e^{-k' \delta } \right)}{k'},
\end{eqnarray}
and
\begin{eqnarray}
%\small
%\label{eq:y1}
%y_1(t_f)&=& y_1(t_0) e^{-k' \delta } + \frac{\left( y(t_0) k' \omega + \bar{I} \right) \left( 1 - e^{-k' \delta } \right) }{k'} \nonumber \\
%&+& \frac{\bar{I} \omega \left( k' \delta  -1 + e^{-k' \delta } \right) }{k'},\\
\label{eq:y2}
y_2(t_f)&=& y_2(t_0)e^{-k' \delta } + y(t_0)(1-\omega)\left(1-e^{-k' \delta }\right)\nonumber \\
&+& \frac{\bar{I} (1-\omega) \left( k' \delta  -1+e^{-k' \delta } \right)}{k'},
\end{eqnarray}
%\end{subequations}
where $k'=k_F / \left( \omega (1-\omega) \right)$, $\delta = t_f - t_0$ and $y=y_1+y_2$. 

\begin{figure}[t]
\includegraphics[width=230pt]{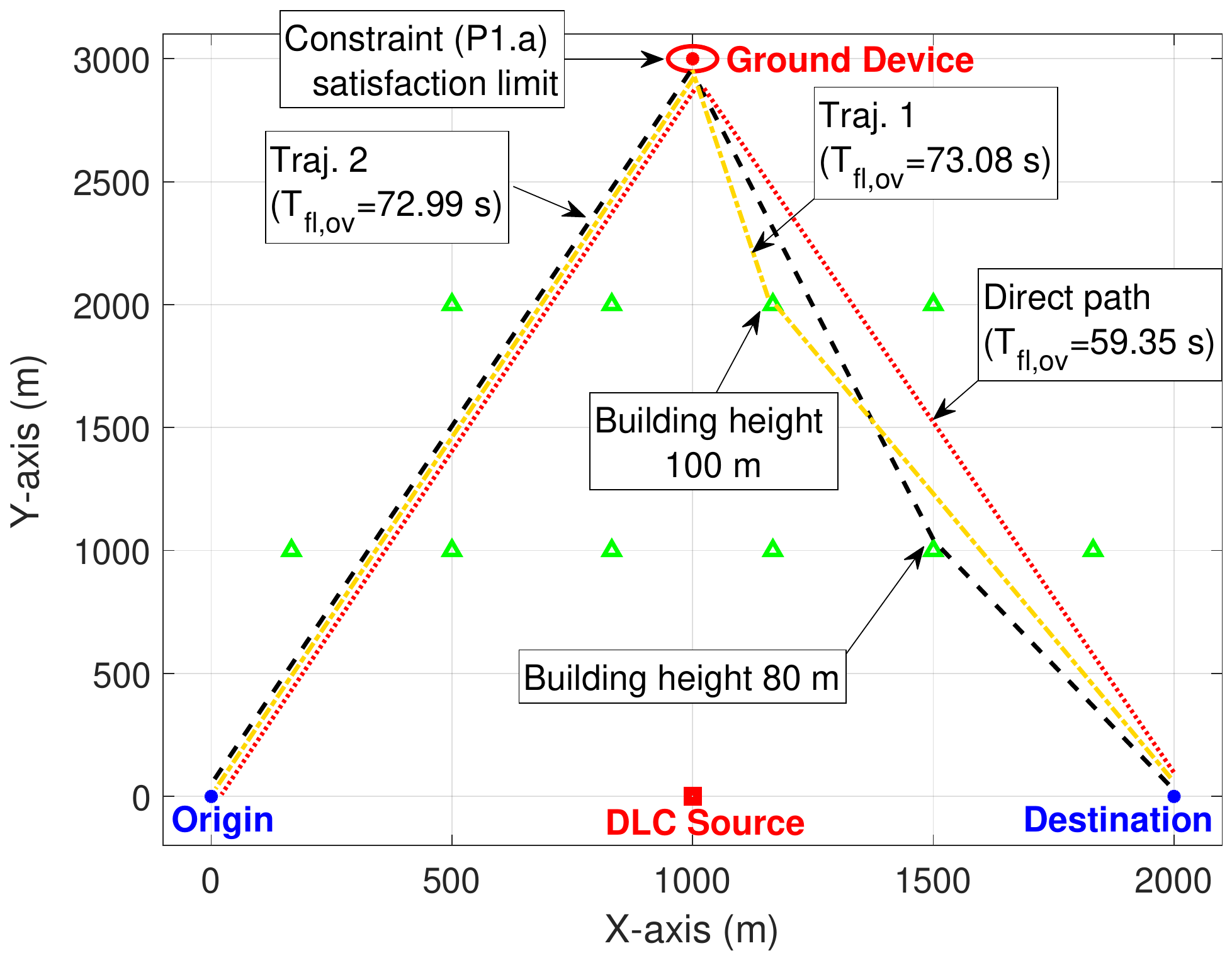}
   \caption{System model and UAV trajectories.}
    \label{Fig1c}
\end{figure}
%\vspace{-5pt}
\section{Revisited Path Planning Problem}
We consider a downlink communication system, consisting of a UAV, a ground device \textcolor{black}{(e.g., a remote server, IoT sink/gateway, etc.)}, a DLC source, \textcolor{black}{and $B=10$ buildings}, placed in the 3D space, \textcolor{black}{where the device is outside the suburban} environment, as shown in Fig. \ref{Fig1c}. The UAV flies from an origin location to a destination during $T\leq T_{\rm{max}}$ time slots, with $T_{\rm{max}}$ is the maximal tolerated flight duration. Among these $T$ time slots, the UAV has to hover and successfully communicate with the ground device for a maximized number of $\Delta$ time slots, where the communication outage probability $P_{\rm{out}}$ has to be kept below a threshold $\varepsilon$. Given a Rician communication channel\footnote{\textcolor{black}{The Rician channel is an accurate air-to-ground channel model \cite{Azari2018}. Other models can be considered, such as the probabilistic model discussed in \cite{Chandrasekharan2016}.}}, $P_{\rm{out}}$ can be obtained as \cite{Zhou2008}
\begin{equation}
\label{Rice_out}
%\small
P_{\rm{out}}=1-Q_1\left( \sqrt{2 K}, \sqrt{\frac{2 \gamma_{\rm{th}} (1+K) N_0}{P_u d^{-\beta}}} \right) \leq \varepsilon,
\end{equation}
where $\gamma_{\rm{th}}$ is the signal-to-noise-ratio (SNR) threshold, $N_0$ is the noise power, $d$ is the distance between the UAV and the ground device, $\beta$ is the path-loss exponent, \textcolor{black}{$K\geq 0$ is the Rice factor, and $Q_1(\cdot,\cdot)$ is the 1$^{st}$-order Marcum Q-function \cite[eq. 4.33]{Simon2004}.} Also, the UAV can receive power from the DLC source {either when hovering or resting on a building. Thus,} we formulate the following problem (P1): 
\begin{subequations}
	\begin{align}
	%\small
	\max_{\mathbf{W}} & \quad 
	%{\eta}={\prod_{i=1}^T \frac{y_1(i)+y_2(i)}{y_1(i-1)+y_2(i-1)}} \tag{P1} \\
	\Delta \tag{P1} \\
	\label{c1}
	\text{s.t.}\quad & P_{\rm{out}}\leq \varepsilon,\; \forall t \in \Delta,  \nonumber \tag{P1.a} \\
	\label{c20} & {\eta}(T) \geq \eta_{0} \tag{P1.b} \\
	\label{c2} & \mathbf{w}(1)=\mathbf{w}_0,\; \mathbf{w}(T)=\mathbf{w}_F,  \tag{P1.c}\\
	\label{c3} & 0 \leq \Delta \leq T_{\rm{max}},  \tag{P1.d}\\
	\label{c4} & z_{\rm{min}} \leq z \leq z_{\rm{max}},  \tag{P1.e}
	\end{align}
\end{subequations}
\noindent where $\eta(T)$ is the {state-of-charge (SOC)}
%left battery charge 
by the end of the mission time $T$ ($T \leq T_{\max}$). It can be expressed either as
\begin{equation}
    \label{eq:etabatt}
    %\small
    \eta(T)=\eta_1(T)=\frac{y_1(T)+y_2(T)}{B}\; \text{(Battery perspective)},
\end{equation}
where $y_1(T)$ and $y_2(T)$ are the wells levels at the end of time slot $T$, respectively, or as
\begin{equation}
    \label{eq:etaenerg}
    \small
    \eta(T)=\eta_2(T)=1-\frac{E_{\rm fl, ov}+E_{\rm hv,ov}-E_{\rm harv}}{E_0} \; \text{(Energy perspective)},
\end{equation}
where $E_{\rm fl, ov}$, $E_{\rm hv,ov}$, and $E_{\rm harv}$ are the consumed overall flying energy, hovering+communication energy, and harvested energy during times $T_{\rm fl,ov}$, $\Delta$, and $T_{\rm harv}$, respectively, such that $T=T_{\rm fl,ov}+T_{\rm harv} \leq T_{\max}$\footnote{To be noted that $T_{\rm harv}$ includes the time spent for hovering, $\Delta$, and the time spent resting on one or several buildings $T_{\rm harv}'=T_{\rm harv}-\Delta$.}. The energy expressions can be obtained using (\ref{eq:Etrav})--(\ref{eq:EEH}). Also, $E_0$ is the nominal initial battery energy, $\varepsilon \in [0,1]$ is the outage probability threshold, $\eta_0$ is the desired SOC level at the end of $T$, $\mathbf{w}(t)=[x(t),y(t),z(t)]$, $t=1,\ldots, T$ is the UAV 3D trajectory, such that $\mathbf{W}=[\mathbf{w}(t)]_{t=1,\ldots,T}$, $\mathbf{w}_0$ and $\mathbf{w}_F$ are the origin and destination 3D-locations respectively, and $z_{\rm{min}}$ and $z_{\rm{max}}$ are the minimum and maximum flying altitudes.
Constraint (\ref{c1}) ensures that the communication is successful between the hovering UAV and ground device. (\ref{c20}) guarantees that the SOC is above a fixed value, while (\ref{c2}) defines the origin and destination locations. Finally, (\ref{c3}) and (\ref{c4}) limit the hovering time and flying altitude, respectively.   
Since $\Delta \leq T_{\max}-T_{\rm fl,ov}-T'_{\rm harv}$, the problem can be rewritten as
\begin{subequations}
	\begin{align}
	%\small
	\min_{\mathbf{W},\Delta} & \quad 
	T_{\rm fl,ov}+T'_{\rm harv} \tag{P2} \\
	\label{c01}
	\text{s.t.}\quad & P_{\rm{out}}\leq \varepsilon,\; \forall t \in \Delta,  \nonumber \tag{P2.a} \\
	\label{c020} & {\eta}(T_{\rm fl,ov}+\Delta+T'_{\rm harv}) \geq \eta_{0} \tag{P2.b} \\
	\label{c02} & \mathbf{w}(1)=\mathbf{w}_0,\; \mathbf{w}(T_{\rm fl,ov}+\Delta+T'_{\rm harv})=\mathbf{w}_F,  \tag{P2.c}
	%\label{c03} & 0 \leq \Delta \leq T_{\max}-T_{\rm fl,ov}-T'_{\rm harv},  \tag{P2.d}\\
	%\label{c04} & z_{\rm{min}} \leq z \leq z_{\rm{max}}.  \tag{P2.e}
	\end{align}
\end{subequations}
\begin{subequations}
	\begin{align}
	%\small
	%\min_{\mathbf{W},\Delta} & \quad 
	%T_{\rm fl,ov}+T'_{\rm harv} \tag{P2} \\
	%\label{c01}
	%\text{s.t.}\quad & P_{\rm{out}}\leq \varepsilon,\; \forall t \in \Delta,  \nonumber \tag{P2.a} \\
	%\label{c020} & {\eta}(T_{\rm fl,ov}+\Delta+T'_{\rm harv}) \geq \eta_{0} \tag{P2.b} \\
	%\label{c02} & \mathbf{w}(1)=\mathbf{w}_0,\; \mathbf{w}(T_{\rm fl,ov}+\Delta+T'_{\rm harv})=\mathbf{w}_F,  \tag{P2.c}\\
	\label{c03} & 0 \leq \Delta \leq T_{\max}-T_{\rm fl,ov}-T'_{\rm harv},  \tag{P2.d}\\
	\label{c04} & z_{\rm{min}} \leq z \leq z_{\rm{max}}.  \tag{P2.e}
	\end{align}
\end{subequations}
For the sake of simplicity, we assume that the UAV either do not rest on a building, i.e., $T'_{\rm harv}=0$, or it rests exactly on one building only, $T'_{\rm harv}>0$\footnote{This assumption is accurate since resting over several buildings would generate longer trajectories, hence more flight energy is consumed at the expense of recharging.}. 

In the first case ($T'_{\rm harv}=0$), minimizing $T_{\rm fl,ov}$ is equivalent to adopting the direct flight trajectory $\textbf{w}_0 \xrightarrow{} \textbf{w}_U \xrightarrow{} \textbf{w}_F$, where $\textbf{w}_U$ corresponds to the furthest UAV location from the ground device satisfying (\ref{c01})\footnote{Typically, this corresponds to the lowest UAV altitude $z_{\min}$\cite{Alzenad2018}.}. Given this trajectory, $T_{\rm fl,ov}$ can be calculated, and thus, $\Delta$ can be found by solving problem (P3):
\begin{subequations}
	\begin{align}
	%\small
	\max & \quad 
	\Delta \tag{P3} \\
	\label{cc1}
	\text{s.t.}\quad & {\eta}(T_{\rm fl,ov}+\Delta) \geq \eta_{0} \tag{P3.a} \\
	\label{cc2} & 0 \leq \Delta \leq T_{\max}-T_{\rm fl,ov} \tag{P3.b}.
	\end{align}
\end{subequations}
The optimal solution is simply obtained for the largest $\Delta  \leq T_{\max}-T_{\rm fl,ov}$ satisfying (\ref{cc1}).

In the second case ($T'_{\rm harv}>0$), the flight time in (P2), $T_{\rm fl,ov}$, can be minimized similarly to the first case by following the most direct trajectory ($\textbf{w}_0 \xrightarrow{} \textbf{w}_U \xrightarrow{} \textbf{w}_b \xrightarrow{} \textbf{w}_F$) or ($\textbf{w}_0 \xrightarrow{} \textbf{w}_b \xrightarrow{} \textbf{w}_U  \xrightarrow{} \textbf{w}_F$), where $\textbf{w}_b$ corresponds to the UAV location above building $b$, $\forall b=1,\ldots,B$. Hence, based on graph theory, we build the potential trajectories and deduce the associated $T_{\rm fl,ov}$, given that the edges' costs correspond to the flight duration between two nodes (i.e., UAV locations) in the graph. Consequently, the problem reduces to the following one: 
\begin{subequations}
	\begin{align}
	%\small
	\max_{T'_{\rm harv}> 0} & \quad 
	\Delta \tag{P4} \\
	\label{ccc1}
	\text{s.t.}\quad & {\eta}(T_{\rm fl,ov}+\Delta+T'_{\rm harv}) \geq \eta_{0} \tag{P4.a} \\
	\label{ccc2} & 0 \leq \Delta \leq T_{\max}-T_{\rm fl,ov}-T'_{\rm harv} \tag{P4.b}.
	\end{align}
\end{subequations}
Given selected building $b$, the problem can be solved using Algorithm \ref{Algo2}, presented next, where $\textbf{0}_{M \times N}$ designates the all zeros matrix of size $M \times N$. In Algorithm \ref{Algo2}, both $T'_{\rm harv}$ and $\Delta$ are varied until the best combination is found. Subsequently, the global solution is obtained by combining the two cases of $T'_{\rm harv}=0$ and $T'_{\rm harv}>0$, as described in Algorithm \ref{Algo3}.

{The proposed Algorithm \ref{Algo3} provides an offline path planning solution, which is adequate for clear spaces, such as rural environments and high altitude operations.  However, it cannot be directly exploited for low altitude dynamic urban environments, as the latter have both static and dynamic obstacles, which could collide with the UAV. In such a case, online (i.e., on-the-fly) or hybrid offline/online methods can be leveraged for more accurate routing. It is expected that the latter impact the UAV energy consumption/harvesting, and hence online replanning is required at the UAV, as discussed in \cite{Xue2014}\nocite{Liu2016_2}--\cite{Yin2018}. The investigation of such approaches from the battery perspective is left for future work.}

\section{Simulation Results}

\begin{algorithm}[t]
%\footnotesize
{
\caption{Hovering and Resting Times Optimization \\Algorithm, given building $b$}
\label{Algo2}
\begin{algorithmic}[1]
\State Initialize $\Delta_{\rm ov}=\textbf{0}_{(T_{\max}-T_{\rm fl,ov})\times (T_{\max}-T_{\rm fl,ov})}$ 
\State Initialize $T'_{\rm harv,ov}=\textbf{0}_{(T_{\max}-T_{\rm fl,ov})\times (T_{\max}-T_{\rm fl,ov})}$ 
\For {$T'_{\rm harv}=1$ to $T_{\max}-T_{\rm fl,ov}$} 
\For{$\Delta=1$ to $T_{\max}-T_{\rm fl,ov}-T'_{\rm harv}$}
\State Calculate $a=\eta(T_{\rm fl,ov}+\Delta+T'_{\rm harv})$
\If {$a \geq \eta_0$}
\State $\Delta_{\rm ov}(T'_{\rm harv},\Delta)=\Delta$
\State $T'_{\rm harv, ov}(T'_{\rm harv},\Delta)=T'_{\rm harv}$
\EndIf
\EndFor
\EndFor
\State Return $\Delta_{\rm opt}=\max (\Delta_{\rm ov})$ \State Return $T'_{\rm opt}=T'_{\rm harv,ov}(\arg \max \Delta_{\rm ov})$
%\State Select best solution among the solutions of $\mathcal{G}_i$ ($i=1,\ldots,n$) graphs.
\end{algorithmic}}
\end{algorithm}

\begin{algorithm}[t]
%\small
{
\caption{Global Optimization Algorithm}
\label{Algo3}
\begin{algorithmic}[1]
\State Solve (P3) and return the solution $\Delta_1$
\State Initialize $\Delta_B=\textbf{0}_{1 \times B}$ and $T'_B=\textbf{0}_{1 \times B}$
\For{$b=1$ to $B$} 
\State Execute Algorithm \ref{Algo2}
\State $\Delta_B(b)=\Delta_{\rm opt}$
\State $T'_B(b)=T'_{\rm opt}$
\EndFor
\State $\Delta_2=\max (\Delta_{B})$ and $T'_2=T'_B(\arg \max \Delta_B)$
\If{$\Delta_1>\Delta_2$}
\State Return $\bar{\Delta}=\Delta_1$
\Else
\State Return $\bar{\Delta}=\Delta_2$ and $\bar{T}=T'_2$
\EndIf
\end{algorithmic}}
\end{algorithm}

%To demonstrate the relevance of the KiBaM battery model in solving the path planning problem, we compare the performance from both energy and battery perspectives, for two different trajectory solutions, namely, the optimal solution obtained using Algorithm \ref{Algo3} and called ``Optimal", and a modified nearest neighbor method called ``NN" (it should be noted that NN is the best solution when the trajectory protocol requires stopping by the closest building either before or after serving the ground device). 

The simulation parameters are set as follows: $\mathbf{w}_0=[0,0,\textcolor{black}{50}]$ m, $\mathbf{w}_F=[\textcolor{black}{2000,0,50}]$ m, DLC source location $\mathbf{w}_s=\textcolor{black}{[1000,0,50]}$ m, and ground device location $\mathbf{w}_u=[1000,\textcolor{black}{3000},0]$ m, \textcolor{black}{whereas the buildings have heights in $\{60,80,100\}$ m, ordered such that buildings with lowest heights are the closest to x-axis edges}. When hovering, the UAV communicates with the ground device during $\Delta$ seconds, where the duration of a time slot is one second. For the sake of simplicity, the UAV experiences only gravity, i.e., external force $\textbf{F}_e = [0, 0, -12.74]$ N, unless stated otherwise. 
For the quadrotor UAV characteristics, we rely on the model of \cite{Morbidi2016}, where the UAV is powered by two 3-cell (3S) LiPo 11.1 V batteries with capacities $B_1$=$B_2$=36000 As (10Ah) \cite{Recharge}, corresponding to $E_0=2 \times(36000 \times 11.1)=799200$ J. The use of two independent batteries allows to \textcolor{black}{extend the battery life by alternatively using one for motion and the other for recharging, and vice versa}. Moreover, we assume that communication channel path-loss $\beta=2$, \textcolor{black}{K-factor $K=20$ (i.e., strong LoS)}, $N_0=10^{-4}$, $\gamma_{th}=-11$ dB, $\varepsilon=10^{-2}$, $z_{\rm{min}}=50$ m, and $z_{\rm{max}}=100$ m. Also, we assume that the duration of 1 time slot is equal to 1 second. The remaining UAV parameters are summarized in Table I. 

%KiBaM model to the combined Peukert's dischage law + direct recharge \cite{DOERFFEL2006}, called ``Peukert", where the consumed battery capacity and the recharge capacity are estimated by $C_{\rm{cons}}=I_{\rm{dis}}^{\zeta} t$ and $C_{\rm{ch}}=I_{\rm{ch}} t$, respectively, and $\zeta=1.3$ is the Peukert constant.} 
%\textcolor{black}{We compare them also to two conservative variations, where the mission time is reduced by 1 min by precaution.}

%the battery is considered empty at $20\%$ of its capacity, namely ``Peukert cons." and ``KiBaM cons.".

%the proposed trajectory, designated ``Prop. GA", to the following benchmarks: 1) ``Bench. 1": Being a user-centric approach, the UAV flies in a straight line to the closest location that allows to satisfy (\ref{c1}). Then, it flies to its destination in a direct line. 2) ``Bench. 2": The UAV flies directly to the closest location located on the segment drawn by the ground user and DLC source locations. 

%After the duration $\Delta$ and if enough time is left, it stops at a close building to DLC source before continuing to destination.        

\begin{table}[t]
    \caption{UAV Parameters \cite{Zhang2018,Morbidi2016,Jongerden2009}}
    \label{tab1} 
\small 
{
    \centering
    \begin{tabular}{lll}
    \hline
       $\varrho$=3.8305$\cdot 10^{-6}$ N/rad/s   & $v_{\rm{max}}$=\textcolor{black}{1060} rad/s & $m$=1.3 Kg  \\
       %\hline
       ($a_1, b_1$)=\textcolor{black}{(0.34,-1.1)} & $\lambda$=\textcolor{black}{1550} nm & $P_u$=\textcolor{black}{5} W \\
       %\hline
        ($a_2, b_2$)=\textcolor{black}{(0.5434, -0.2761)}   & $T_f$=0.04 N.m  & $R$=0.2 $\Omega$  \\ 
        %\hline
        $\kappa_0$=2.2518$\cdot 10^{-8}$ N.m/rad/s & $\kappa_V$= 920 rpm/V  &  $\omega$=0.8 \\
        %\hline
        $\kappa_E$=$\kappa_T$=9.5493/$\kappa_V$  & \textcolor{black}{$e_{nom}= 11.1$ V} & $I_{\rm{ch}}$=10 A  \\
        %\hline
        $D_f$=2 $\cdot 10^{-4}$ N.m.s/rad & \textcolor{black}{$\alpha$=0.1019}  & $e_{\rm{tr}}$=1 V \\
        % \hline
        $J$=4.1904 $\cdot 10^{-5}$ Kg.m$^2$ & $k_F$=4.5 $\cdot 10^{-5}$ &  \\
         \hline
    \end{tabular}
    }
\end{table}

%\begin{table}[t]
%\small{
%\begin{center}
%\caption {UAV Parameters %\cite{Zhang2018,Morbidi2016,Jongerden2009}}
% \begin{tabular}{|l|l| } 
% \hline
% $\varrho$=3.8305$\cdot 10^{-6}$ N/rad/s  &  $m$=1.3 Kg   \\ [0.5ex]  \hline
%  $v_{\rm{max}}$=\textcolor{black}{1060} rad/s  & $P_u$=\textcolor{black}{5} W   \\ %1047.197
% \hline
%  ($a_1, b_1$)=\textcolor{black}{(0.34,-1.1)}  & \textcolor{black}{$\alpha=0.1019$ (clear air)}   \\
% \hline
%($a_2, b_2$)=\textcolor{black}{(0.5434, -0.2761)}  & $\lambda$=\textcolor{black}{1550} nm   \\ 
% \hline
% $\kappa_V$= 920 rpm/V  & $R$=0.2 $\Omega$  \\
% \hline
% $\kappa_E=\kappa_T$=9.5493/$\kappa_V$  & $\kappa_0=2.2518 \cdot 10^{-8}$ N.m/rad/s \\
% \hline
% $k_F=4.5 \cdot 10^{-5}$ min$^{-1}$  &  $T_f$=0.04 N.m  \\
% \hline
% $J=4.1904 \cdot 10^{-5}$ Kg.m$^2$  &  $\omega$=0.8  \\
% \hline
% $D_f$=0.0002 N.m.s/rad  & $I_{\rm{ch}}$=10 A \\
% \hline
% $e_{\rm{nom}}=3 \times 3.7= 11.1$ V & $e_{\rm{tr}}$=1 V \\
% \hline
%\end{tabular}
%\end{center}}
%\label{tab1} 
%\end{table}

\begin{table}[t]
    \label{TabDirect}
    \begin{center}
    %\small 
    %\small
        \caption{Performances of different perspectives}
    \begin{tabular}{|c|c|c|c|}
    \hline
        \makecell{Approach} & $\Delta$ (s) & $T$ (s) & $\eta(T)$ (\%)    \\ \hline \hline
        \makecell{Direct path\\($T_{\max}$=800 s)\\(Battery Persp.)} & \textbf{740.65} & 800 & \makecell{\textbf{$\eta_1(T)$=24.27}\\ $\eta_2(T)$=-12.79 \\ $\eta_3(T)$=24.27} \\ \hline
        \makecell{Direct path\\($T_{\max}$=800 s)\\(Ener. Persp.)} & 697 &  756.35 & \makecell{$\eta_1(T)$=28.31\\ \textbf{$\eta_2(T)$=5.01} \\ $\eta_3(T)$=28.30} \\ \hline
        \makecell{Direct path\\($T_{\max}$=800 s)\\(Adj. Ener. Persp.)} & \textbf{740.65} &  800 & \makecell{{$\eta_1(T)$=24.27}\\ $\eta_2(T)$=-12.79 \\ \textbf{$\eta_3(T)$=24.27}} \\ \hline \hline
        \makecell{Direct path\\($T_{\max}$=1200 s)\\(Battery Persp.)} & \textbf{950} &  1009.35 & \makecell{\textbf{$\eta_1(T)$=5.09}\\ $\eta_2(T)$=-25.58 \\ $\eta_3(T)$=11.71} \\ \hline
        \makecell{Direct path\\($T_{\max}$=1200 s)\\(Ener. Persp.)} & 697 &  756.35 & \makecell{$\eta_1(T)$=28.31\\ \textbf{$\eta_2(T)$=5.01} \\ $\eta_3(T)$=33.17} \\ \hline
        \makecell{Direct path\\($T_{\max}$=1200 s)\\(Adj. Ener. Persp.)} & \textbf{950} &  1009.35 & \makecell{{$\eta_1(T)$=5.09}\\ $\eta_2(T)$=-25.58 \\ \textbf{$\eta_3(T)$=5.09}} \\ \hline
    \end{tabular}
    \end{center}
    \label{tab:my_label}
\end{table}

\begin{figure}[t]
	\centering
	\includegraphics[width=265pt]{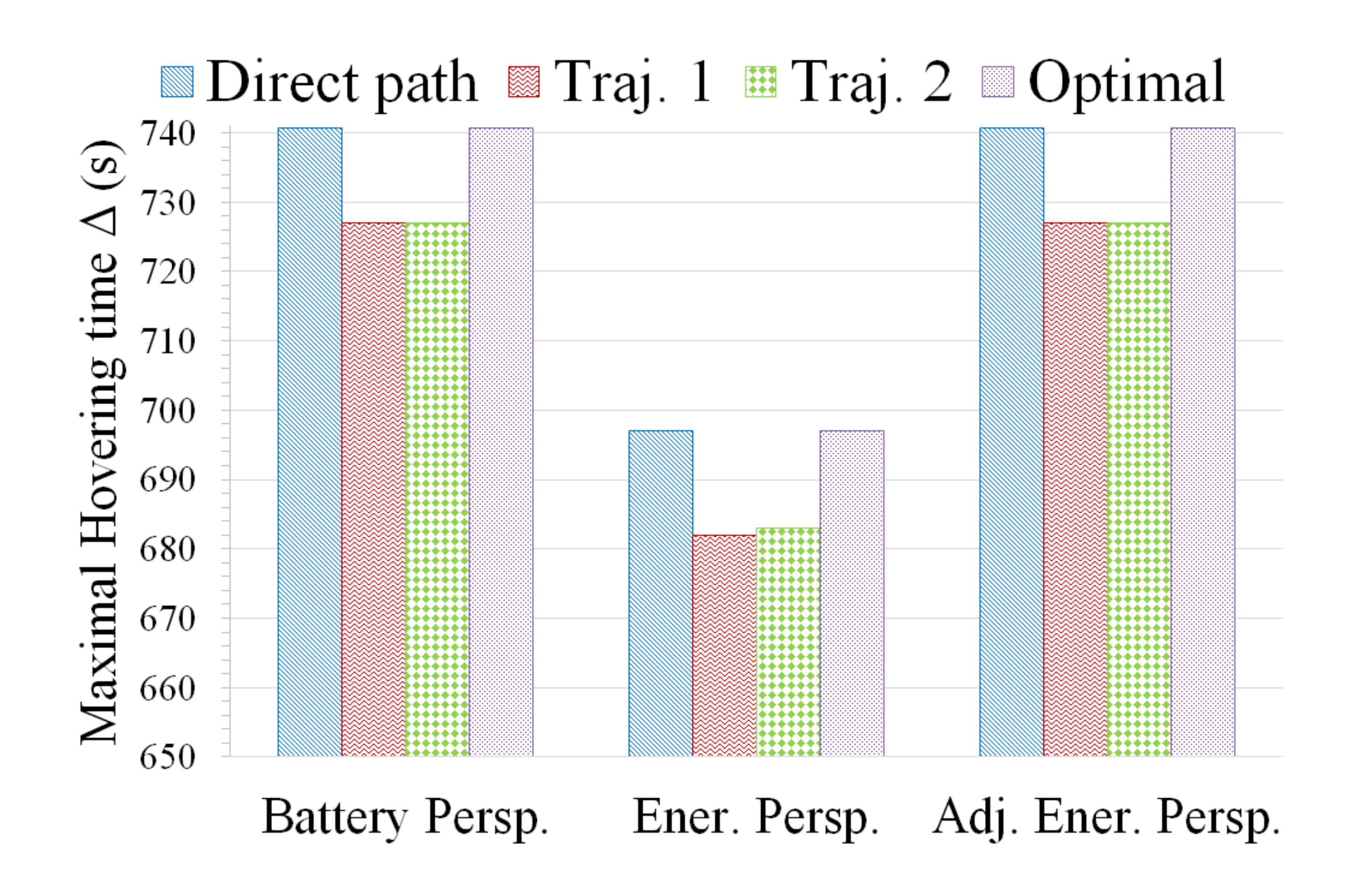}
	\caption{Maximal hovering time vs. resolution perspective ($T_{\max}$=800 seconds).}
	\label{Fig11}
\end{figure}
\begin{figure}[t]
	\centering
	\includegraphics[width=265pt]{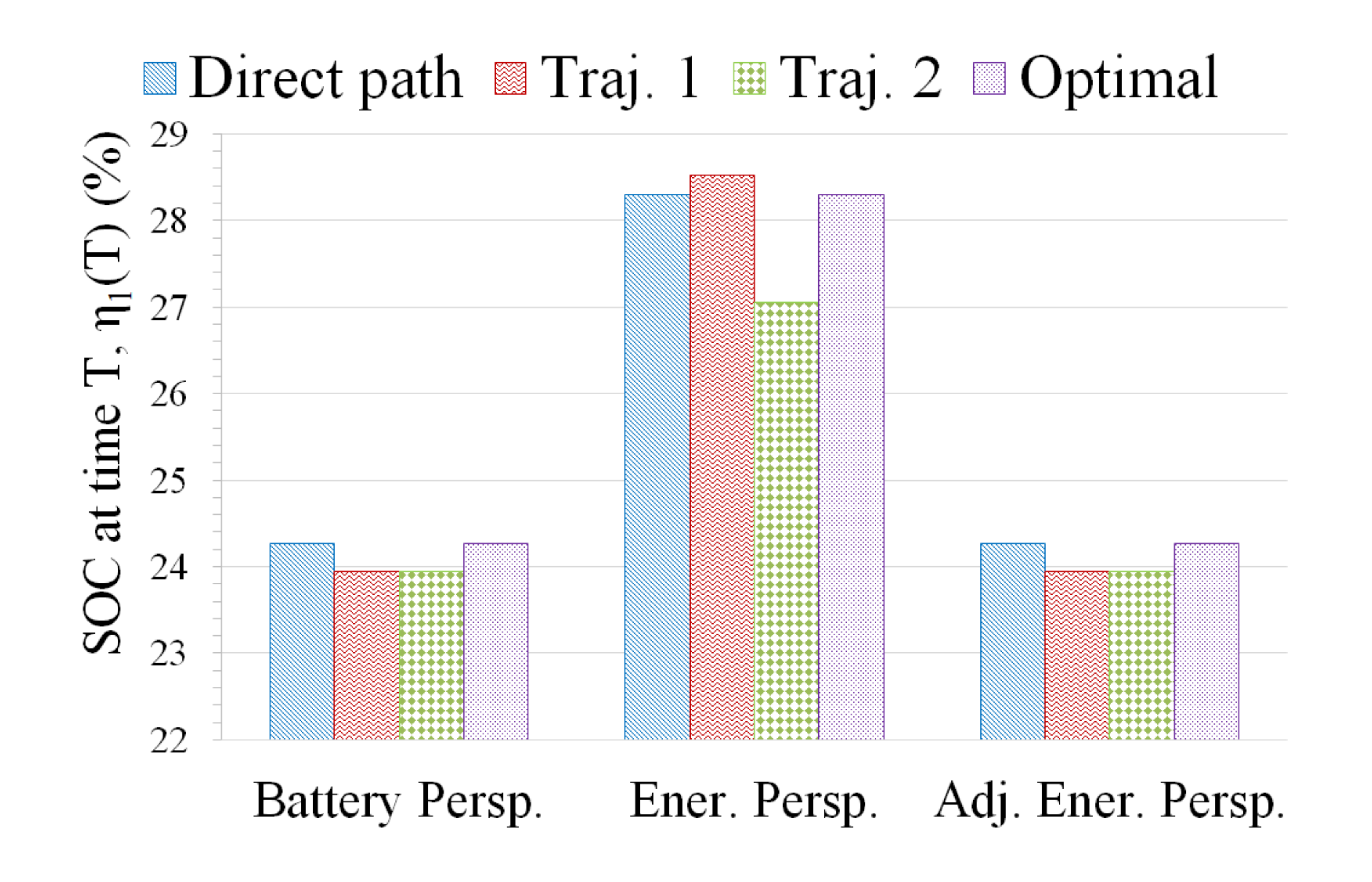}
	\caption{SOC at time $T$ vs. resolution perspective ($T_{\max}$=800 seconds).}
	\label{Fig12}
\end{figure}

%\begin{table}[t]
%\label{tab2} 
%\small{
%\begin{center}
%\caption {Performances}
% \begin{tabular}{|l | l |l |l | l | } 
% \hline
% Approach & $B_1$(T)As & $B_2$(T)As & $\eta$(\%) & T(sec)  \\
% \hline
%  Bench. 1 & 33837 & 31444.6 & 90.66 & 94.7  \\
% \hline
%Bench. 2 (w/o stop.) & 33862.8 & 31441.8 & 90.7 & 95  \\ 
% \hline
% Bench. 2 (w/ stop.) & 33572 & 31650.6 & 90.6 & 119.1  \\
% \hline
% Prop. GA & 34924.6 & 33828.3 & 95.5 & 119  \\
% \hline
%\end{tabular}
%\end{center}}
%\end{table}

{First, we evaluate in Table II the solutions to (P3) given the direct path ($\textbf{w}_0 \xrightarrow{} \textbf{w}_U \xrightarrow{} \textbf{w}_F$), and for different $T_{\max}$ and perspectives, i.e., condition (\ref{cc1}) is expressed either using (\ref{eq:etabatt}) or (\ref{eq:etaenerg}). For $T_{\max}$=800 s, the battery perspective, called ``Battery Persp.", achieves better $\Delta$ than the energy perspective, named ``Ener. Persp.". Indeed, the latter is overestimating the energy consumption (due to energy calculation using $e_{nom}$), and thus provides a lower $\Delta$=697 s. In order to adjust this perspective, we propose to modify $\eta_2(T)$ as
\begin{equation}
\eta_3(T)=1-\frac{E_{\rm fl,ov}}{E_1}+\frac{E_{\rm hv,ov}}{E_2}-\frac{E_{\rm harv}}{E_0},    
\end{equation}
where $E_1=2 \times (36000 \times 15.4)= 1108800$ J and $E_2=2 \times (36000 \times 14.33)= 1032036$ J are the initial battery capacities calculated using $e_r(t)$ in (\ref{eq:DCmotor}), which correspond to $e_r(t)=15.4$V for flying and $e_r(t)=14.33$V for hovering.
The division of $E_{\rm fl,ov}$ and $E_{\rm hv,ov}$ by $E_1$ and $E_2$ respectively is justified by the different flight regime/battery usage compared to resting/harvesting. As it can be seen, the adjusted energy perspective, denoted ``Adj. Ener. Persp.", achieves the same $\Delta$ as for ``Battery Persp.". For $T_{\max}$=1200 s, similar results are obtained, where both ``Battery Persp." and ``Adj. Ener. Persp." achieve the best performance, with a mission time $T$=1009.35 s, which is below $T_{\max}$.   
%\textcolor{red}{the ``Adj. Ener. Persp." is showing the best $\Delta$=1026.5, however, this approach cannot be relied on since the associated $\eta_1$=-2.17$<$0, i.e., practically the UAV would fail from the sky as its battery is emptied before reaching the destination. Consequently, the ``Battery Persp." is the most reliable one, providing $\Delta$=950.}
}

{In Figs. \ref{Fig11}--\ref{Fig12}, we solve (P1) with different trajectory approaches and illustrate the resulting maximum hovering time, $\Delta$, and the SOC from the battery perspective, $\eta_1(T)$, given that $T_{\max}$=800 s. (P1) is solved from the ``Battery Persp.", ``Ener. Persp.", and ``Adj. Ener. Persp." as defined in the previous paragraph.
We define ``Direct path", ``Traj. 1", ``Traj. 2" and ``Optimal" trajectory approaches, where ``Direct path" is described in the previous paragraph, ``Traj. 1" corresponds to a modified nearest neighbor approach with the UAV passing by or resting over the closest building to the ground device, ``Traj. 2" is defined as the shortest trajectory with the UAV passing by or resting over one building, and ``Optimal" is obtained using Algorithm \ref{Algo3}. These trajectories are depicted in Fig. \ref{Fig1c}. In Fig. \ref{Fig11}, ``Direct path" provides the maximal hovering time $\Delta$ for all solution perspectives as it follows the shortest path. Also, both ``Battery Persp." and ``Adj. Ener. Persp." achieve better performances than ``Ener. Persp.". According to Fig. \ref{Fig12}, the SOC measured from the battery perspective, $\eta_1(T)$, respects the constraint $\eta_0$, with ``Ener. Persp." saving a lot of the battery's capacity due to its low hovering time. Therefore, the ``Ener. Persp." has a conservative approach to the problem.}

\begin{figure}[t]
	\centering
	\includegraphics[width=267pt]{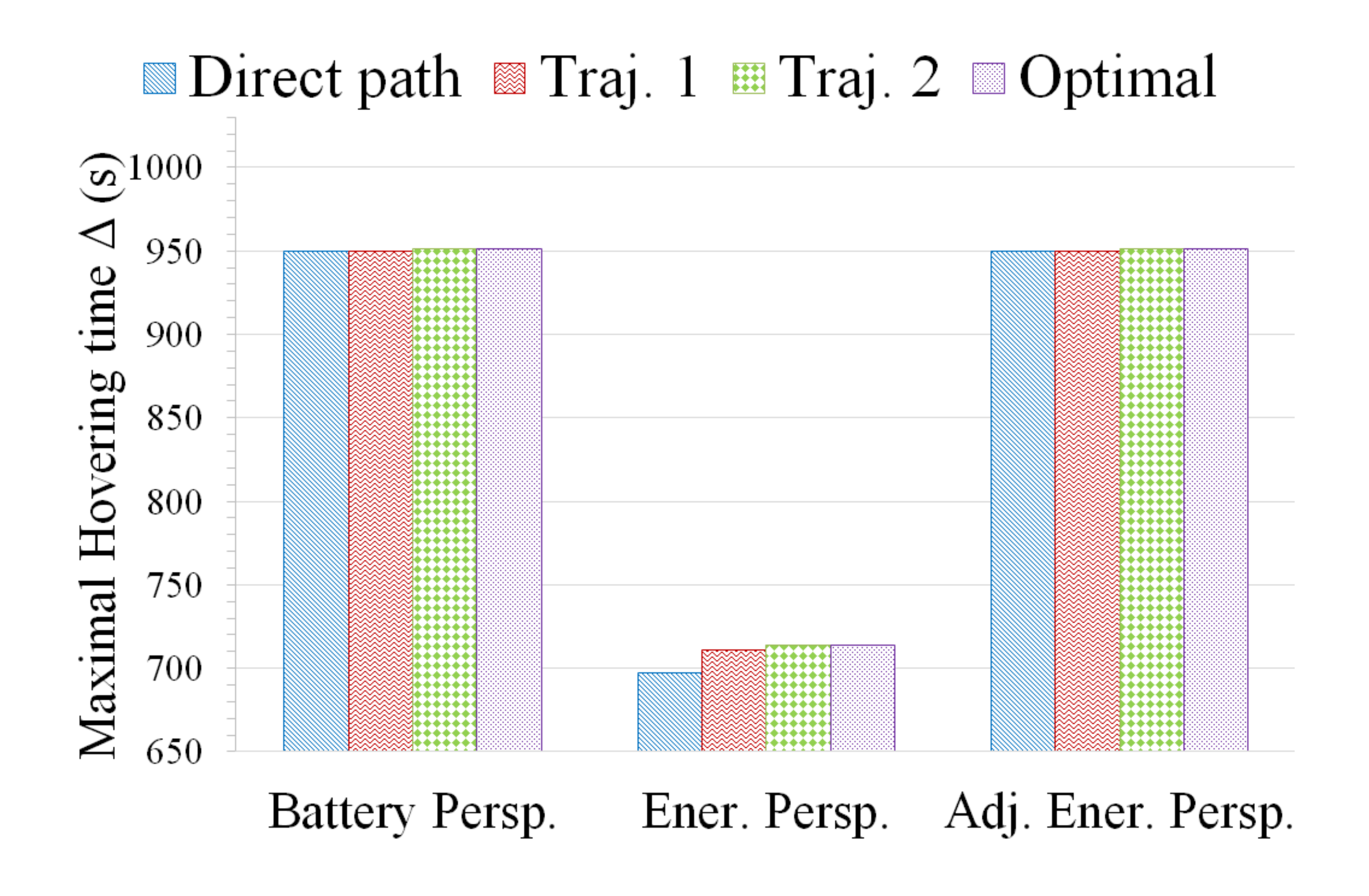}
	\caption{Maximal hovering time vs. resolution perspective ($T_{\max}$=1200 seconds)}
	\label{Fig21}
\end{figure}

\begin{figure}[t]
	\centering
	\includegraphics[angle=-90,width=1.01\linewidth]{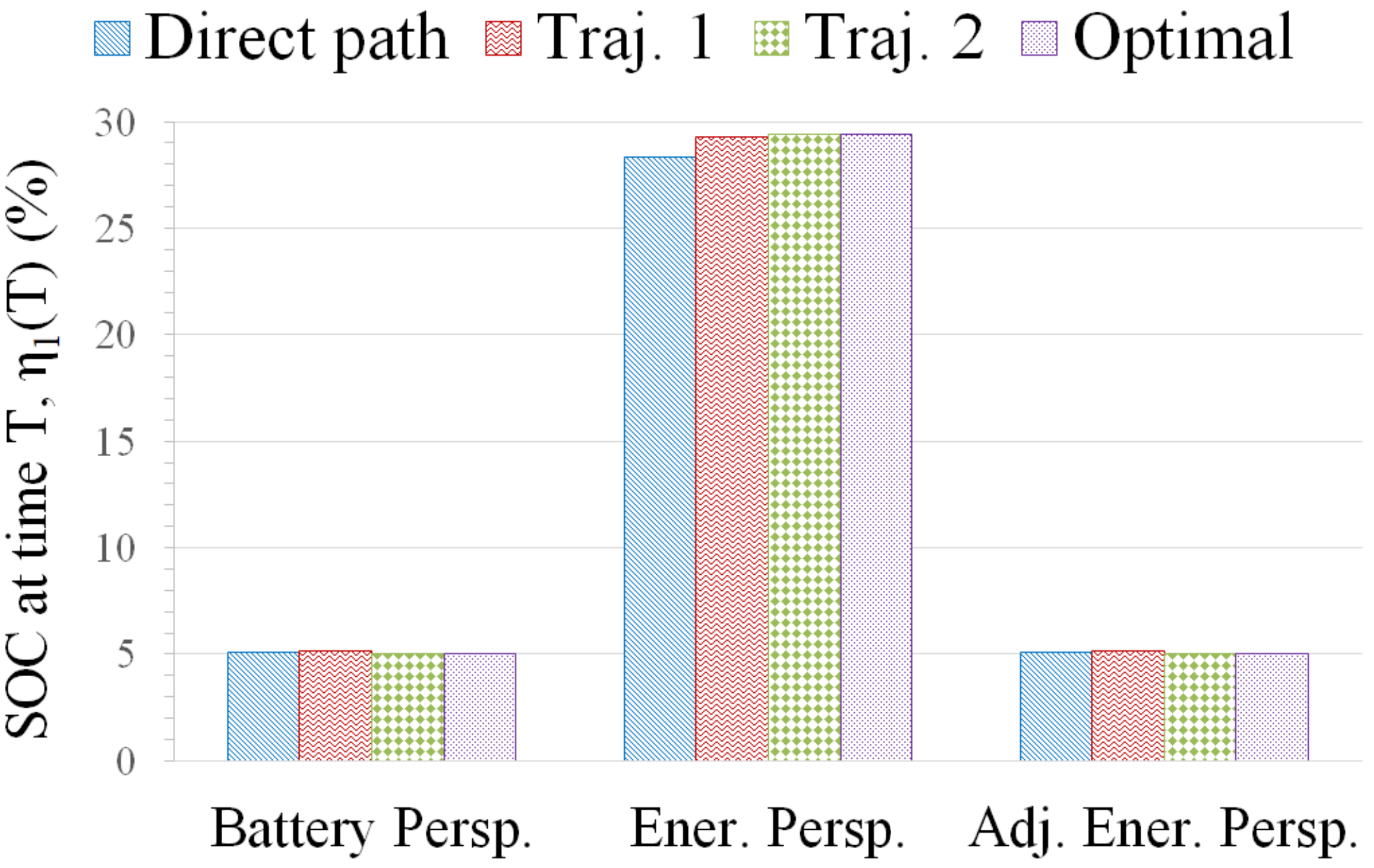}
	\caption{SOC at time $T$ vs. resolution perspective ($T_{\max}$=1200 seconds).}
	\label{Fig22}
\end{figure}

{In Figs. \ref{Fig21}--\ref{Fig22}, the same performances are depicted, but for $T_{\max}$=1200 s. In Fig. \ref{Fig21}, ``Traj. 2" slightly outperforms the other approaches, which is also the optimal solution. Indeed, since $T_{\max}$ is high, the UAV is capable of hovering for a longer time and compensates for the extra consumed energy by resting over one building and recharging its battery. 
%Although $\eta_1(T)\geq \eta_0$ for ``Battery Persp." and ``Ener. Persp.", it is below 0 for the ``Adj. Ener. Persp.". Indeed, the ``Adj. Ener. Persp." ensures that $\eta_3(T)\geq \eta_0$, which provides very high $\Delta$. This consumes additional battery capacity that exceeds the battery's limit, and causes the UAV's failure. Consequently, ``Adj. Ener. Persp." can be seen as a risky approach when $T_{\max}$ is high. 
Moreover, both ``Battery Persp." and ``Adj. Ener. Persp." have the same best performance. Although these perspectives are similar, the ``Battery Persp." is recommended when designing energy-efficient UAV-based missions, as it simplifies the SOC expression in contrast to ``Adj. Ener. Persp.", which requires different initial battery calculations, depending on the UAV's motion regimes. 
}


\begin{figure}[t]
	\centering
	\includegraphics[width=0.65\linewidth,angle=-90]{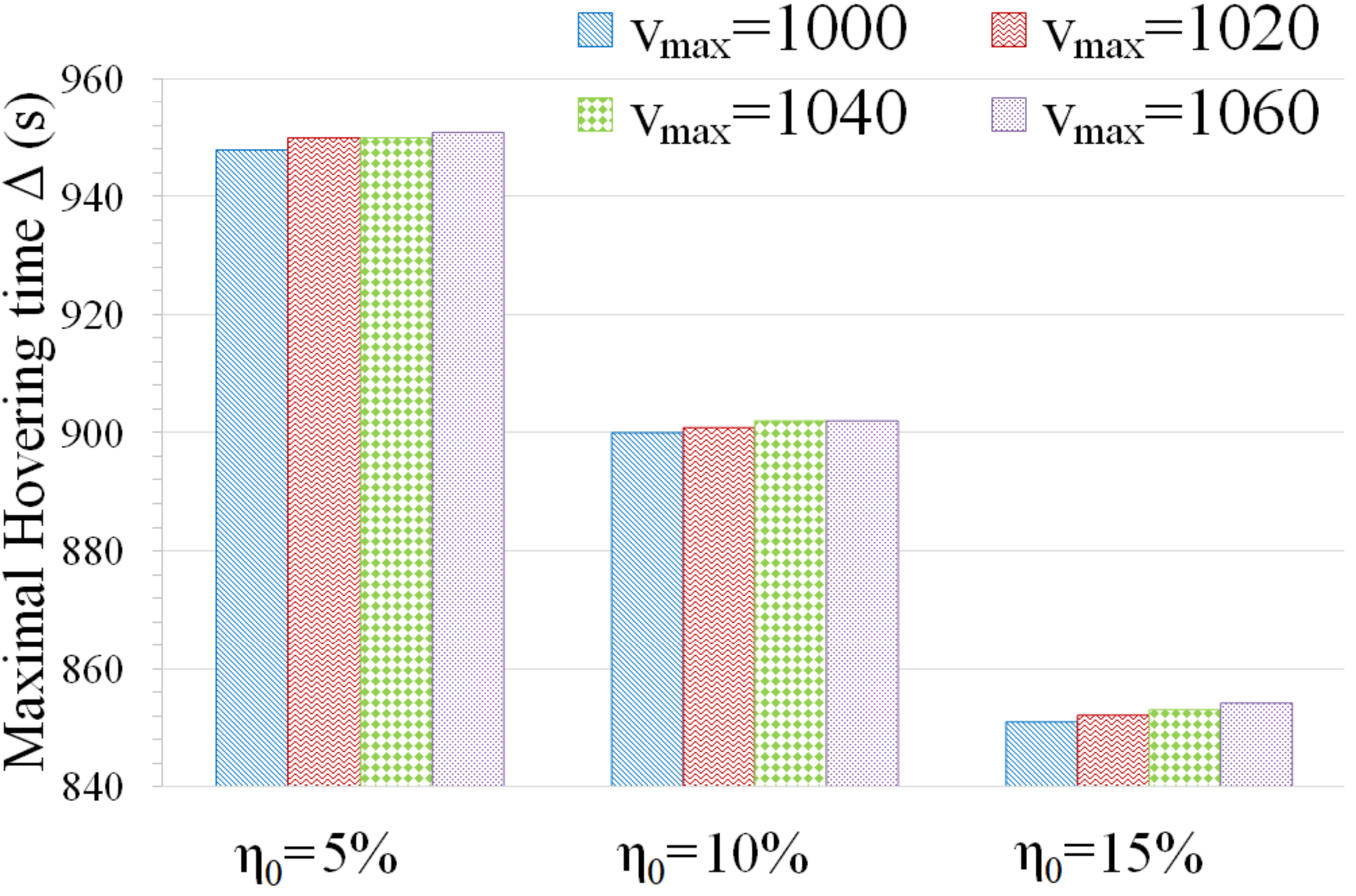}
	\caption{{Maximal hovering time vs. the desired SOC level $\eta_0$ ($T_{\max}$=1200 seconds).}}
	\label{Fig3new}
\end{figure}

{Fig. \ref{Fig3new} investigates the impact of the maximal rotor velocity $v_{\rm{max}}$ (rad/s) for different desired SOC levels $\eta_0$. The optimal trajectory is determined using the ``Battery Persp.'', given $T_{\max}$=1200 s.
For any $\eta_0$, the maximal hovering time $\Delta$ improves with $v_{\max}$. Indeed, the UAV moves at a higher speed, thus reducing the flying time and extending its hovering time. Although more energy is consumed for flying at higher speeds, the latter is largely compensated by the additional time spent for energy harvesting. Moreover, as $\eta_0$ becomes more stringent from 5$\%$ to 10$\%$ then 15$\%$, $\Delta$ decreases by approximately 5.25$\%$ and 10.5$\%$, respectively. This performance reduction is required to satisfy (\ref{ccc1}).}

\begin{figure}[t]
    \centering
    \includegraphics[width=220pt]{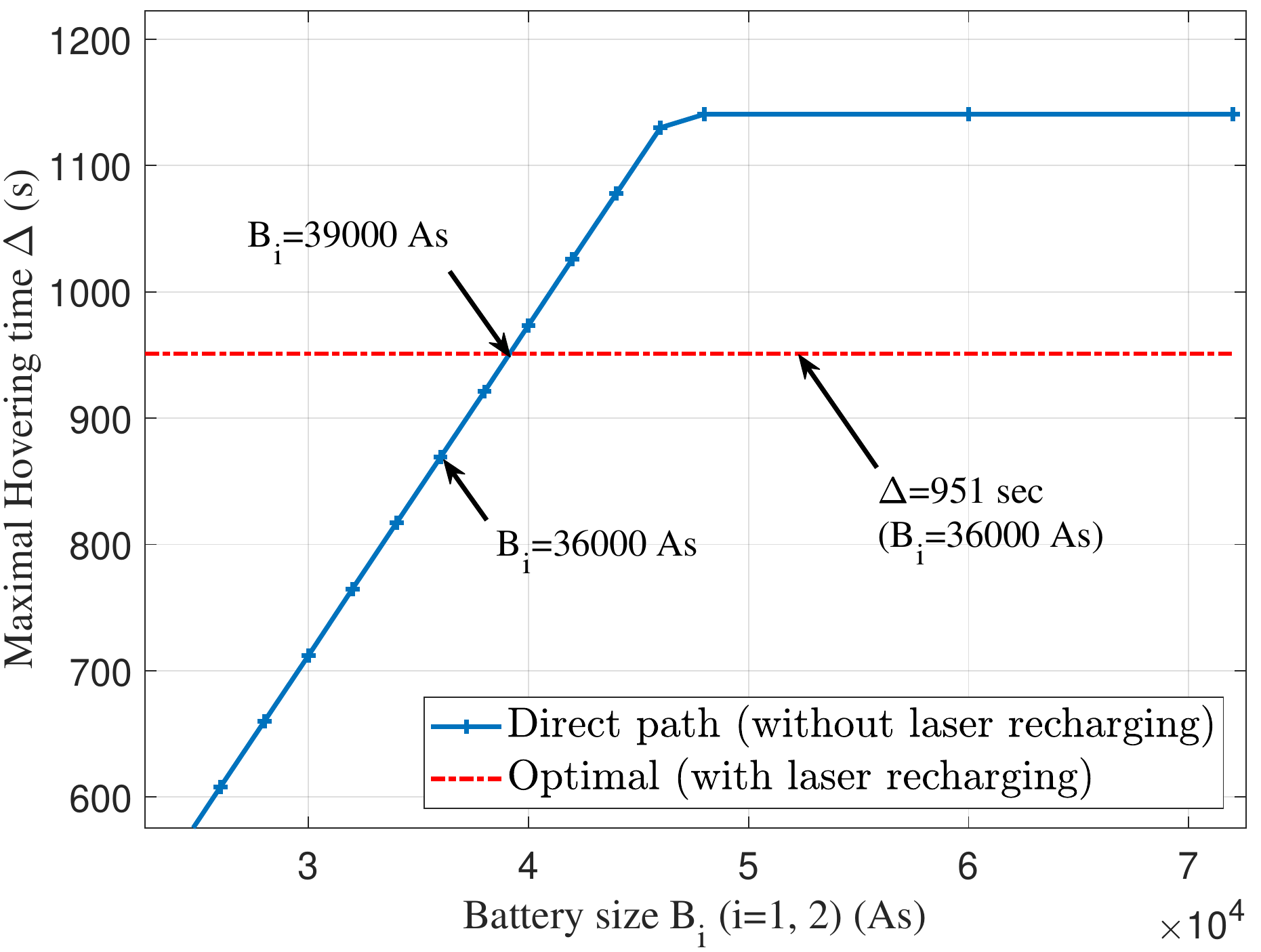}
\caption{{Maximal hovering time vs. battery size $B_i$ ($i=1,2$) ($T_{\max}=1200$ seconds).}}
    \label{Fig4compare}
\end{figure}

{In order to illustrate the advantageous use of laser recharging, we evaluate in Fig. \ref{Fig4compare} the maximal hovering time performance of the optimal solution (for the previously described system), to the direct path that does not support laser recharging, but can instead use a different battery size. As it can be seen, when $B_i<39000$ As ($i=1,2$), the optimal solution with laser recharging outperforms the direct path. For $\Delta$=951 s, supporting laser recharging saves $(39000-36000)=3000$ As, i.e., 8.33$\%$, in battery size. When $B_i$ exceeds $39000$ As, the direct path provides a better performance due to the additional power in its large-sized battery. In any case, the latter's performance saturates at $B_i=48000$ As since $T_{\max}$ cannot be exceeded.}

\begin{figure}[t]
    \centering
    \includegraphics[width=213pt]{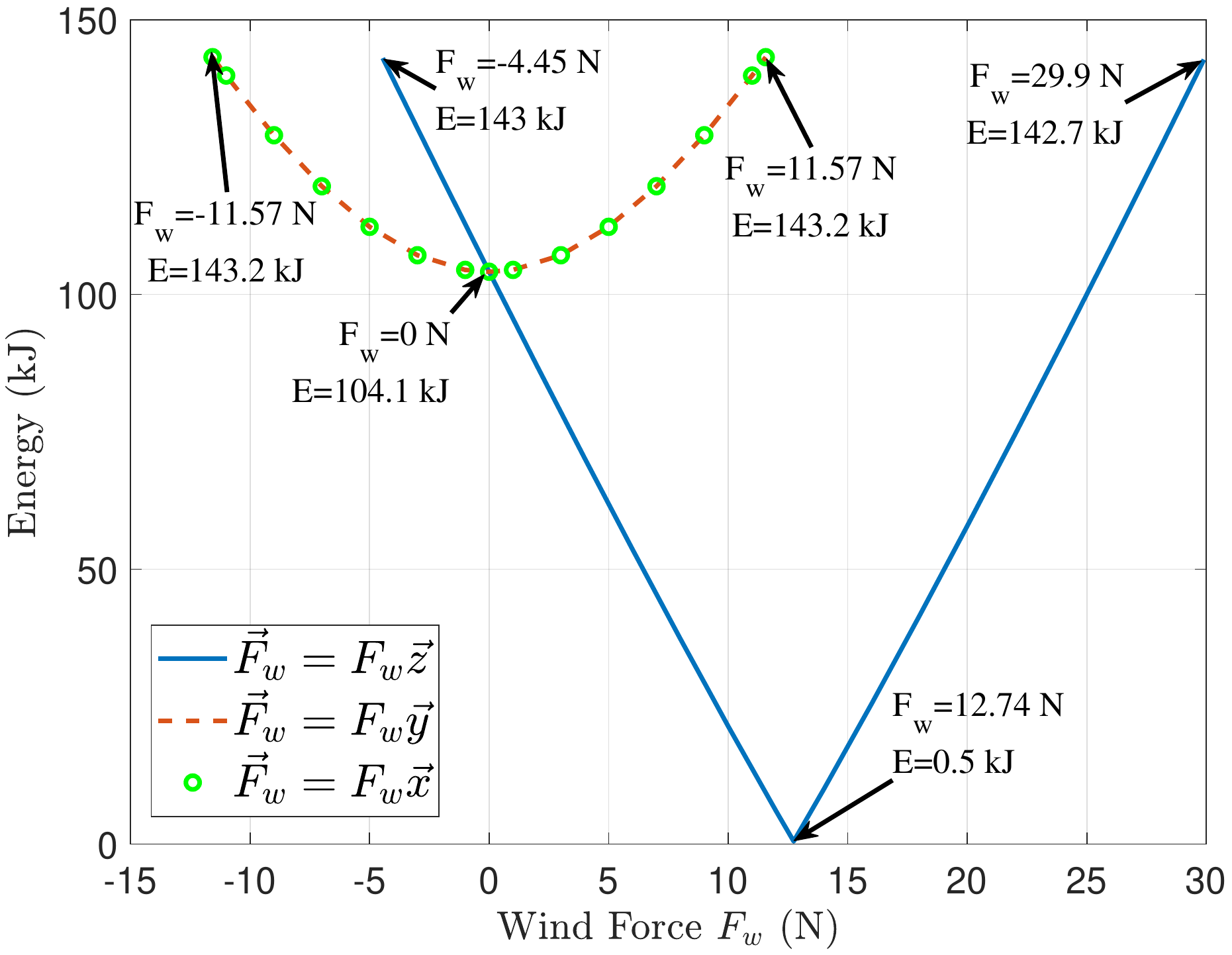}
\caption{\textcolor{black}{Hovering+communication energy $E_{\rm hv,ov}$  vs. wind force.}}
    \label{Fig3c}
\end{figure}

\begin{figure}[t]
    \centering
    \includegraphics[width=225pt]{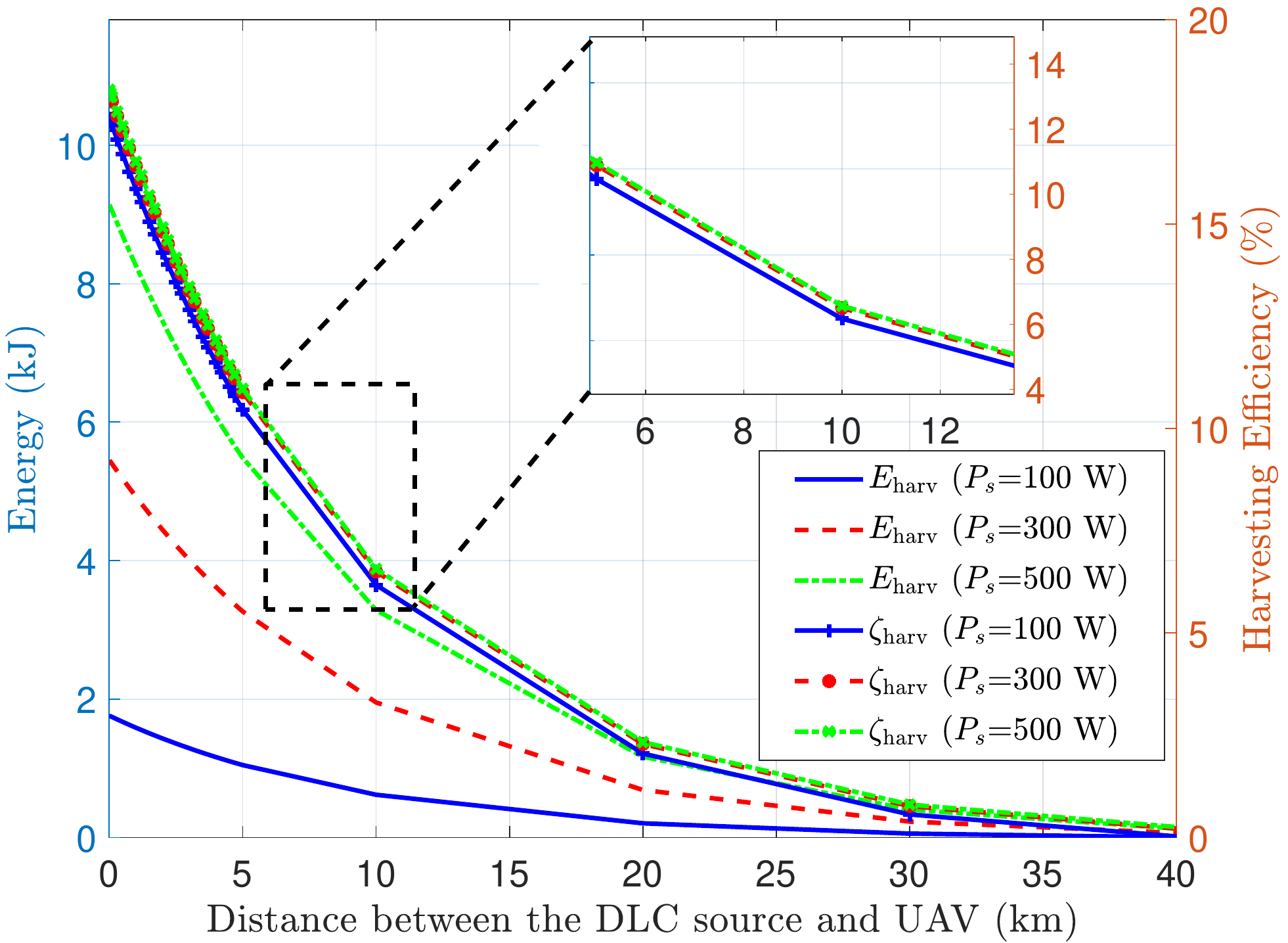}
	\caption{Harvested energy and efficiency vs. distance DLC source-UAV.}
	\label{Fig4c}
\end{figure}

Let $\vec{F}_w=\vec{F}_e - m \vec{g}$ be the turbulence (e.g., wind) force that hits the UAV when hovering during $\Delta=\textcolor{black}{100}$ s to serve the ground device, where $m \vec{g}$ is the gravity. The impact of $F_w$ is investigated in Fig. \ref{Fig3c}. For $\vec{F}_w=F_w \vec{x}$ (resp. $F_w \vec{y}$), $E_{\rm hv, ov}$ has a parabolic shape, where the lowest consumed energy is for $F_w=0$N. Also, the curves are bounded by minimum and maximum $F_w$ values $-11.57$ N and $11.57$ N, corresponding to the maximum wind force that can be handled by the UAV without losing its balance. Going beyond these values requires a higher $v_{\rm{max}}$ and hence higher power.
Along $\vec{z}$, \textcolor{black}{$F_w \in [-4.45, 12.74]$N} counters gravity, hence the UAV can hover using less angular velocity $v_r$ ($r=1,\ldots,4$) and power, which decreases $E_{\rm hv,ov}$. However, for $F_w \in [12.74,\textcolor{black}{29.9}]$N, the wind pushes the UAV to provide more power in order to stay aloft. Beyond these values, the UAV would lose its vertical balance. %Unlike the previous case, $B(\Delta)$ evolves linearly with $F_w \vec{z}$.   

Fig. \ref{Fig4c} evaluates $E_{\rm{harv}}$ (eq.(\ref{eq:EEH})) and the harvesting efficiency $\zeta=\frac{P_0}{P_s}$ when the UAV is hovering for \textcolor{black}{100} s, as functions of the distance between the DLC source and the UAV, and for different $P_s$. As the distance increases, both $E_{\rm{harv}}$ and $\zeta$ degrade due to path-loss. \textcolor{black}{For distances below 1 km, $\zeta$ is around 17.5\%. $E_{\rm{harv}}$ can be significantly improved by increasing $P_s$.}  
%It is to be noted that while $\zeta$ is around \textcolor{black}{18}\% for distances below 300 m, we found through simulations that these performances drop in the range 6\%--15\% when the UAV is moving from a location to another.

%However, we notice that $E_{\rm{harv}} \leq 3$kJ, which is an order of magnitude below $E_{\rm{trav}}$ and $E_{\rm{hov}}$. 

%\begin{figure}[t]
%	\centering
%	\includegraphics[width=160pt]{Fig5_crop2.pdf}
%	\caption{Consumed/Harvested battery ratio vs. battery size.}
%	\label{Fig5c}
%\end{figure}

%In Fig. \ref{Fig5c}, we illustrate the consumed/harvested battery ratio, defined as $\gamma=1-\frac{y_1(t_f)+y_2(t_f)}{y_1(t_0)+y_2(t_0)}$ and calculated using (\ref{eq:well1})-(\ref{eq:well2}). As $B_i$ ($i=1,2$) increases, $\gamma$ decreases, since the amount of consumed/harvested energy is the same. Also, hovering+communication consumes more energy than traveling, dominated by the hovering energy. Whereas, WPT compensates for some of the lost energy. For instance, a gain of $2\%$ of $B_1=10800$As is achieved using WPT.        

%As the current systems demonstrate a harvesting efficiency below $25\%$,   

\section{Conclusion}
In this paper, we established the relationship between power, energy, and battery dynamics for a laser-charged quadrotor UAV. Based on the obtained expressions, we revisited a path planning problem and solved it optimally. 
The obtained results emphasize the importance of the battery perspective when investigating UAV-based challenges, such as path planning, UAV placement, and resource optimization. Indeed, it is shown that the conventional energy perspective is very conservative and does not exploit optimally the available energy. Nevertheless, it can be adjusted by adequately evaluating the energy as a function of the UAV motion regime. Finally, the impact of {several parameters, including UAV velocity, desired SOC level, battery size,} turbulence {strength}, and distance to recharging source on the UAV's {performances} is studied, which highlights {the relevance of these parameters in designing UAV missions, 
}
%the influence of the lateral and vertical turbulence, 
as well as the importance of the laser source for {battery size savings and} efficient recharging. {The validation of these results through a real-environment test-bed will be conducted as a future work.}  

%In this letter, we proposed a simple quadrotor UAV model, where energy and battery dynamics are investigated. By leveraging the motors' and battery electrical models, we derived closed-form expressions of consumed/harvested energy and battery levels, and these were illustrated through an experiment. These results will be of great interest to researchers working on future energy-efficient aerial networks. 

%\vspace{-13.5pt}

% or



%\section*{Acknowledgment}

\ifCLASSOPTIONcaptionsoff
  \newpage
\fi



%\bibliographystyle{IEEEtran}

%\balance

\bibliographystyle{IEEEtran}
\bibliography{IEEEabrv,tau}

\end{document}